\documentclass[fleqn,usenatbib,onecolumn]{mnras}
\usepackage{newtxtext,newtxmath}
\usepackage{url}

\usepackage[T1]{fontenc}
\DeclareRobustCommand{\VAN}[3]{#2}
\let\VANthebibliography\thebibliography
\def\thebibliography{\DeclareRobustCommand{\VAN}[3]{##3}\VANthebibliography}

\usepackage{graphicx}	
\usepackage{amsmath}	

\title[3d 21-cm lensing]{Three-dimensional weak gravitational lensing of the 21-cm radiation background}
\author[J. A. Lozano Torres, B.M. Sch{\"a}fer]{
Jose Agustin Lozano Torres \thanks{E-mail: jalozanotorres@gmail.com}, Bj{\"o}rn Malte Sch{\"a}fer
\\
 Astronomisches Rechen-Institut, Zentrum für Astronomie der Universität Heidelberg, Philosophenweg 12, Heidelberg D-69120, Germany
}

\date{Accepted XXX. Received YYY; in original form ZZZ}
\pubyear{2022}

\begin{document}
\label{firstpage}
\pagerange{\pageref{firstpage}--\pageref{lastpage}}
\maketitle

\begin{abstract}
We study weak gravitational lensing by the cosmic large-scale structure of the 21-cm radiation background in the 3d-weak lensing formalism. The interplay between source distance measured at finite resolution, visibility and lensing terms is analysed in detail and the resulting total covariance $C_{\ell}(k,k')$ is derived. The effect of lensing correlates different multipoles through convolution, breaking the statistical homogeneity of the 21-cm radiation background. This homogeneity breaking can be exploited to reconstruct the lensing field $\hat{\phi}_{\ell m}(\kappa)$ and noise lensing reconstruction $N_{\ell}^{\hat{\phi}}$ by means of quadratic estimators. The effects related to the actual measurement process (redshift precision and visibility terms) change drastically the values of the off-diagonal terms of the total covariance $C_{\ell}(k,k')$. It is expected that the detection of lensing effects on a 21-cm radiation background will require sensitive studies and high-resolution observations by future low-frequency radio arrays such as the SKA survey.
\end{abstract}


\section{Introduction}
The dark ages and the transition to a reionised Universe is a central topic in cosmology, in particular with large-scale experiments mapping out the matter distribution at these high redshifts through the spin-flip transition of neutral hydrogen at 21 cm. The fluctuations in the brightness of this 21-cm transition can be resolved spatially and provides a unique window at structure formation at high redshifts in its dependence on fundamental physics. The detection of the 21-cm signal from high redshifts in radio interferometry is among the main objectives of low frequency radio arrays like Mileura Widefield Array (MWA) \citep{2005ASPC..345..452M, 972, Bowman_2007}, the Primeval Structure Telescope (PAST) \citep{doi:10.1142/S0217732304014288} and the Low Frequency Array (LOFAR) \citep{LOFAR, 10.1111/j.1745-3933.2005.00048.x}. One expects a feeble 21-cm signal in comparison to a high quantity of foreground emission coming from atomic processes of astrophysical objects \citep{zahn_lensing_2006, jalivand_intensity_2018}, but the science of the 21-cm background has the potential to test fundamental physics to unprecedented levels, in particular with the future Square Kilometer Array (SKA) \citep{patel_weak_2015, harrison_ska_2016, mellema_reionization_2013, weltman_fundamental_2020, kitching_euclid_2015, jarvis_cosmology_2015}

There is, analogously to other radiation backgrounds, a gravitational lensing effect introduced by the matter distribution at lower redshifts, which is distorting the brightness distribution of the 21-cm background in a characteristic way. As the neutral hydrogen atoms release 21-cm photons at different redshifts across the Universe, their trajectories will be deflected by the gravitational potentials of large scale structures. The weak gravitational lensing presents another source of fluctuations and changes the intrinsic power spectra as well as breaking statistical homogeneity, and is determined by density fluctuations along the photon path between emission and observer. Fortunately, the effects of weak gravitational lensing on the cosmic microwave background (CMB) are understood in detail and the theory for lensing of radiation backgrounds is fully worked out \citep{A, B, PhysRevD.95.043508,Cooray_2002,PhysRevD.98.023535, LEWIS20061, Hu_2001, hollenstein_constraints_2009, cooray_weak_2003, manzotti_super-sample_2014, cooray_lensing_2004, merkel_interplay_2013, schmittfull_joint_2013, cooray_weak_2002, mangilli_optimal_2013, pourtsidou_testing_2015, carbone_lensed_2009, pan_dependence_2014, challinor_geometry_2002,benoit-levy_full-sky_2013, kesden_lensing_2003, lewis_lensed_2005, hanson_cmb_2011, amblard_weak_2004, bucher_cmb_2012, pal_towards_2014,carron_maximum_2017}, and can be readily applied to the 21-cm background: Both background are well described by homogeneous and isotropic random fields with close-to-Gaussian statistics, with an intervening deflection field that is likewise statistically homogeneous and isotropic, with slight deviations from Gaussianity due to nonlinear structure formation. Of course the assumption of Gaussianity is better at higher redshift and on larger scales: Concerning the radiation background, effects of baryonic dynamics and radiative transport \citep{Watkinson:2015vla,Watkinson:2018efd} are affecting statistics in addition to nonlinear structure formation \citep{Shaw:2019qin, Mondal:2016hmf} and can potentially be controlled by machine learning methods \citep{Doussot:2019rdm}, and the deflection field acquires non-Gaussian statistics, which can be handled perturbatively similar to the CMB \citep{merkel_gravitational_2011, mangilli_optimal_2013, bohm_effect_2018}.

The similarity between the CMB and neutral hydrogen radiation background is striking, but in contrast to the CMB, which is a two-dimensional radiation source plane, the 21-cm radiation background is regarded a continuum of source planes covering a wide range in redshift. The study of weak lensing in this scenario can be achieved either by tomography or an alternative approach called 3d weak lensing which introduces spherical harmonics and Bessel functions to take into account the angular and radial decomposition into modes, as outlined in \citet{doi:10.1046/j.1365-8711.2003.06780.x, PhysRevD.72.023516, doi:10.1093/mnras/stv193, refId02}. In this work, we investigate the effects of weak gravitational lensing by the cosmic large-scale structure on the 21-cm radiation background. The 21-cm radiation background emitted from distinct values of redshift $z_{e}$ is lensed by the gravitational potential of the matter distribution located between the emission point and us. Its exploration requires both large and deep areas of the sky, a simultaneous treatment of the spherical sky geometry and of the extended radial coverage. In the scenario of weak lensing, earlier works such as cosmic tomography has been applied, dividing the emission sources into redshift bins and using 2D harmonic analysis but losing radial information as the number of redshift bins decreases. To solve this, some previous works have made one step further extending the all-sky formalism into 3 dimensions to take into account a degree of uncertainty in redshift \citep[e.g.][]{doi:10.1046/j.1365-8711.2003.06780.x}.

Furthermore, \citet{doi:10.1046/j.1365-8711.2003.06780.x} shows that the spherical Fourier-Bessel decomposition is a natural basis for the analysis of fields, for large-scale weak lensing surveys which have distance information at a given uncertainty. In particular, we will adopt the 3d dimensional weak lensing formalism developed in \citet{doi:10.1046/j.1365-8711.2003.06780.x, PhysRevD.72.023516} to derive the expressions such as covariance of the deflection angle, angular power spectrum, the lensing potential in terms of the redshift distribution, and incorporate visibility functions that replace the redshift distribution of sources in weak cosmic shear. We will also derive an expression for the lensing quadratic estimator and noise variance in the spherical Fourier-Bessel picture. The structure of this paper is the following: In Sect.~\ref{sec:21-cm radiation} we summarise the fundamental aspects of the 21-radiation background and give a short introduction into the theory of 3d weak lensing in Sect.~\ref{weak lensing}, leading to the analytical results in Sect.~\ref{analytical expressions}. In section \ref{quadratic estimator} we derive a quadratic lensing estimator from 21-cm radiation background and present our numerical results in Sect.~\ref{results}, which are discussed in Sect.~\ref{conclusions}. Throughout this article, we have used a basic $\Lambda$CDM cosmology with a prior on spatial flatness. Cosmological parameter values have been chosen to coincide with the \citet{Planck}-measurements, i.e. $\Omega_{m0}=0.316$, $ \Omega_{\Lambda 0}=0.684$, $\Omega_{b0}h^{2}=0.0224$, $h=0.674$, $\sigma_{8}=0.811$ and $n_{s}=0.965$.

\section{21 cm radiation background}\label{sec:21-cm radiation}
The spins of the electron and proton in neutral hydrogen can exhibit spin-flip transitions which are visible if the temperature of the spin-system is not in thermal equilibrium with the CMB nor with the ambient gas. The spin-flipping transition emits or absorbs a low-energy photon with a wavelength $\lambda=(1+z)21.106~ \text{cm}$. This 21-cm signal has been taken as a cosmological probe since it can help to trace the matter distribution before the reionisation epoch and follow the transition from neutral to ionised intergalactic gas. However, this signal must be analysed carefully when exploring the Universe at high redshifts. The first problem is that the 21-cm signal becomes very weak to detect it during reionisation period; essentially, the window between CMB-release and reionisation defines a visibility function for the 21cm-background. Another crucial issue is to determine the redshift information of location of 21-cm radiation sources as the volume of survey increases. Incorrect redshift information results in a loss of radial information in the survey as well as in a systematic error, and in this application the redshift uncertainty is essentially given by the receiver bandwidth. In this study, we will model the 21-cm brightness fluctuations as well as the lensing fields as Gaussian random fields, and assume that there is no biasing of the 21-cm field relative to the matter distribution. Details of the biasing model would include a detailed understanding of the thermal evolution as well as of baryonic structure formation, and deviations from Gaussianity in the deflection field are in principle computable in perturbation theory in analogy to non-Gaussian lensing of the CMB.

\subsection{21-cm signal}\label{21-cm signal}
The amplitude of the 21-cm radiation released by the spin-flip of neutral hydrogen is determined by the abundance of neutral hydrogen atoms in the excited stated relative to the ground state which is ruled by the spin temperature, $T_{S}$,
\begin{equation}
\frac{n_{1}}{n_{0}}=\frac{g_{1}}{g_{0}}e^{\left( -\frac{T_{*}}{T_{S}} \right)},
\end{equation}
where $T_{*} = h\nu_{21}/k_{B}\approx 68$ mK, $g_{i}$ is the statistical weight of the energetic state $i$ and $T_{S}$ is the 21-cm spin transition temperature. During the epoch prior to reionisation it is assumed that all the hydrogen is neutral, and fluctuations in the neutral hydrogen density, $\Delta_{HI}$ produce fluctuations in the brightness temperature. Since the fluctuating part of the 21-cm brightness temperature is of interest here, we define that the brightness temperature observed at a frequency $\nu$ along a given direction $\boldsymbol{\hat{n}}$ is given by
\begin{equation}
    T_{b}(\chi_{\nu},\boldsymbol{\hat{n}})=\overline{T}_{b}(z)\Big(\Delta_{HI}(\chi_{\nu},\boldsymbol{\hat{n}})+1\Big),
\end{equation}
where the frequency of observation point is related to the corresponding redshift by $\nu=1420/(1+z)$ MHz. The radial comoving distance at a certain frequency is given by
\begin{equation}
    \chi_{\nu}=\int_{0}^{z}\mathrm{d}z'\frac{c}{H_{0}E(z')}
\end{equation}
where $E(z)=[\Omega_{m}(1+z)^{3}+\Omega_{\Lambda}]^{1/2}$ is the expansion function. The brightness temperature for the 21-cm line is given by \cite{FURLANETTO2006181} 
\begin{equation}
    \overline{T}_{b}(z)\approx 25 \quad \text{mK} \quad \sqrt{\frac{0.15}{\Omega_{m}h^{2}}}\left(\frac{\Omega_{b}h^{2}}{0.022} \right) \left( \frac{1-Y}{0.76} \right)\sqrt{\frac{1+z}{10}},
\end{equation}
where $Y\approx 0.24$ is the helium mass fraction, $\Omega_{b}$ is the average density parameter of baryons today relative to the critical density but it can be approximated as the average density of neutral hydrogen atoms $\Omega_{HI}(z)$. Additionally, $\Delta_{HI}$ represents the neutral hydrogen overdensity evaluated at the coordinates $(\chi_{\nu},\boldsymbol{\hat{n}})$. It should be noticed that we ignored the peculiar velocities of the neutral hydrogen and the expansion of the Universe only contributes to the redshift.

\section{3d weak lensing analysis} \label{weak lensing}

\subsection{Transformation of a scalar field}
The application of 3d weak lensing analysis has been widely used to explore the properties of the large-scale structure with the specific application in mind to constrain dark energy, mathematically analysed in \citet{Leonard,kitching_systematic_2008,kitching_3d_2011,pratten_baos_2013,leistedt_3dex:_2012,grassi_detecting_2014, merkel_intrinsic_2013,leistedt_3d_2015,kitching_3d_2014-1,asorey_recovering_2012,zieser_cross-correlation_2016,merkel_parameter_2017,mancini_3d_2018}. We consider the expansion of an isotropic and homogeneous neutral hydrogen overdensity field $\Delta_{HI}=\Delta (\textbf{r})$ into spherical harmonics and Bessel functions. In a flat geometry, the field can be decomposed in the 3-dimensional spherical Fourier-Bessel basis set $f({\textbf{r}})$ defined here by
\begin{equation}
f_{\ell m}(k)=\sqrt{\frac{2}{\pi}}\int \mathrm{d}\boldsymbol{r} f(\textbf{r}) j_{\ell}(k\chi)Y_{\ell m}^{*}(\boldsymbol{\hat{n}})
\label{transform}
\end{equation}
and its inverse 
\begin{equation}
f(\textbf{r}) = 
\sqrt{\frac{2}{\pi}}\int k^{2}\mathrm{d}k \sum_{\ell=0}^{\infty}\sum_{m=-\ell}^{\ell}f_{\ell m}(k)j_{\ell}(k\chi)Y_{\ell}(\boldsymbol{\hat{n}}),
\label{inversetransform}
\end{equation}  
where $j_{\ell}(k\chi)$ is a spherical Bessel function in the radial direction, $Y_{\ell m}(\boldsymbol{\hat{n}})$ the spherical harmonics on the surface of a unit sphere and $k$ is the radial wave-number. The 3-dimensional spherical Fourier-Bessel power spectrum $C_{\ell}(k)$ of the scalar field $f(\boldsymbol{r})$  is given by the 2-point function of the spherical Fourier-Bessel coefficients $f_{\ell m}(k)$ written as
\begin{equation}
    \langle f_{\ell m}(k) f^{*}_{\ell 'm'}(k')\rangle =C_{\ell }(k)\delta(k-k')\delta^{K}_{\ell \ell '}\delta^{K}_{mm'},
\end{equation}
if the field is statistically isotropic and homogeneous. On the other hand, if we take the radial dependence of the field into account, then the covariance is not longer homogeneous which becomes
\begin{equation}
    \langle f_{\ell m}(k) f^{*}_{\ell 'm'}(k')\rangle=C_{\ell}(k,k')\delta^{K}_{\ell \ell'}\delta^{K}_{mm'}.
    \label{newcovariance}
\end{equation} 
 The covariance \ref{newcovariance} for the 21-cm brightness with all observable effects and ultimately weak lensing incorporated is the primary target of this paper.

\subsection{Observable effects: source visibility and redshift distribution}\label{observable effects integrals}
In general, a cosmological field, such as the neutral hydrogen overdensity field $\Delta(\boldsymbol{r})$, with $\textbf{r}=(\chi,\theta, \phi)$,  will be only partially  observed due to a finite survey volume. In this scenario, we can describe the neutral hydrogen overdensity field as 
\begin{equation}
\Delta(\textbf{r}) = W_{\Delta}(\textbf{s})\delta(\textbf{r}),
\end{equation}
 where $W_{\Delta}(\textbf{s})$ is the visibility function with estimated position $\textbf{s}$, $\delta(\textbf{r})$ is the usual matter overdensity field. So the 3d power spectrum of the coefficients $C_{\ell}(k,k')$ depends on $\ell$, $k$ and $k'$, in contrast to the 3d power spectrum of a statistically homogeneous field, as the neutral hydrogen fraction is changing as a function of redshift and introduces a radial modulation of the brightness. The neutral hydrogen overdensity field $\Delta(\textbf{r})$ is decomposed as
\begin{equation}
\Delta_{\ell m}(k)=\sqrt{\frac{2}{\pi}}\int \mathrm{d}s s^{2} W_{\Delta}(\textbf{s})\delta(\textbf{r})j_{\ell }(ks)\int\mathrm{d}\boldsymbol{\hat{n}}\: Y_{\ell m}^{*}(\theta,\phi). 
\end{equation}
 It is noticed that the coordinates of the comoving radial part from deep surveys are given as a redshift with some uncertainty, involving an estimated radial comoving coordinate $s$ and true radial coordinate $\chi$. These two comoving radial coordinates are related by a conditional probability $p(\chi|s)$ that can be modelled by a Gaussian.
\begin{equation}
p(\chi|s) = \frac{1}{\sqrt{2\pi}\sigma_{z}}\text{exp}\left[- \frac{(z_{s}-z_{\chi})^{2}}{2\sigma_{z}^{2}} \right],
\label{conditionalprobability}
\end{equation}
 where $z_{\chi,s}$ are the redshifts associated with comoving and estimated radial coordinates $\chi$, $s$ and $\sigma_{z}$ is the error which may vary with redshift. Therefore, one can derive harmonics that represent the average value of the expansion coefficients by using the relation established between the estimated distance from the measured redshift, $s$, and the true distance $\chi$ in terms of the conditional probability 
\begin{equation}
\begin{aligned}
\Delta_{\ell m}(k)&=\sqrt{\frac{2}{\pi}}\int \mathrm{d}s s^{2} \int \mathrm{d}\chi p(\chi|s)     W_{\Delta}(\textbf{s})\delta(\textbf{r})j_{\ell }(ks) \int \mathrm{d}\boldsymbol{\hat{n}}Y_{\ell m}^{*}(\theta,\phi).
\label{coefficientdeltalm}
\end{aligned}
\end{equation}
 Furthermore, such a Gaussian error leads to redshift smoothing. With a smooth redshift, $W_{\Delta}(\textbf{s})$ is smoothed, so it can be approximated by the smoothed number density at $\textbf{r}$, $W_{\Delta}^{s}(\textbf{r})$. For deep surveys, we can approximate this by the average number density, divided into a radial part and an angular part
\begin{equation}
W_{\Delta}^{s}(\textbf{r})=\tilde{W}(\chi)M(\theta,\phi),
\end{equation}
 with $M=1$ in survey. If we define the spherical transform of $W_{\Delta}^{s}(\textbf{r})\delta (\textbf{r})$ by $g_{\ell m}(k)$, i.e.
\begin{equation}
W_{\Delta}^{s}(\textbf{r})\delta(\textbf{r})=\sqrt{\frac{2}{\pi}}\sum_{\ell =0}^{\infty}\sum_{m=-\ell }^{\ell } \int \mathrm{d}k k^{2} g_{\ell m}(k)j_{\ell }(k\chi)Y_{\ell m}(\theta,\phi)
\end{equation}
Hence, equation~\ref{coefficientdeltalm} can be written
\begin{equation}
\begin{aligned}
\Delta_{\ell m}(k)&=\frac{2}{\pi}\int \mathrm{d}s s^{2} \int \mathrm{d}\chi p(\chi|s) \sum_{\ell '=0}^{\infty}\sum_{m'=-\ell '}^{\ell '} \int \mathrm{d}k'k'^{2}g_{\ell 'm'}(k')j_{\ell '}(k'\chi)j_{\ell }(ks) \int \mathrm{d}\boldsymbol{\hat{n}} Y_{\ell 'm'}(\theta,\phi)Y_{\ell m}^{*}(\theta, \phi).
\label{delta2}
\end{aligned}
\end{equation}
 We can simplify equation \ref{delta2} by using the orthogonality of the spherical harmonics $\int \mathrm{d}\boldsymbol{\hat{n}} Y_{\ell 'm'}(\theta,\phi)Y_{\ell m}^{*}(\theta,\phi)=\delta_{\ell \ell '}^{K}\delta_{mm'}^{K} $ leading to
\begin{equation}
\Delta_{\ell m}(k)=\int \mathrm{d}k' k'^{2} Z_{\ell }(k,k')g_{\ell m}(k')
\end{equation}
 where $Z_{\ell }(k,k')=\frac{2}{\pi}\int \mathrm{d}s s^{2}\int \mathrm{d}\chi p(\chi|s)j_{\ell }(k'\chi)j_{\ell }(ks) $ and  $g_{\ell m}(k)$ may be calculated by direct substitution of the expansion of $\delta$, yielding
\begin{equation}
g_{\ell m}(k)=\int \mathrm{d}k'k'^{2}M_{\ell }(k,k')\delta_{\ell m}(k')
\end{equation} 
 where $M_{\ell }(k,k')=\frac{2}{\pi}\int \mathrm{d}\chi \chi^{2}j_{\ell }(k'\chi)j_{\ell }(k\chi)\tilde{W}(\chi) $. So we have reached a final expression for the spherical Fourier-Bessel coefficient of the neutral hydrogen overdensity field written as 
\begin{equation}
\Delta_{\ell m}(k)=\int \mathrm{d}k' k'^{2} Z_{\ell }(k,k')\int \mathrm{d}k'' k''^{2} M_{\ell }(k',k'')\delta_{\ell m}(k'')
\label{coefficient}
\end{equation} 
 Consequently, the covariance can be expressed as 
\begin{equation}
\begin{aligned}
C_{\ell }^{\Delta\Delta}(k,k')= & \int \mathrm{d}k_{1}k_{1}^{2}\int \mathrm{d}k_{2}k_{2}^{2}\int \mathrm{d}k_{3}k_{3}^{2}\int \mathrm{d}k_{4} k_{4}^{2}  Z_{\ell }(k,k_{1})Z_{\ell '}(k',k_{2}) M_{\ell }(k_{1},k_{3})M_{\ell '}(k_{2},k_{4}) \frac{P^{\delta \delta}_{\ell }(k_{3})}{k_{3}^{2}}\delta^{1D}(k_{3}-k_{4})\delta_{\ell \ell '}^{K}\delta_{mm'}^{K}.
\end{aligned}
\end{equation} 
 By using the relation $  \langle \delta_{\ell m}(k_{3})\delta_{\ell 'm'}(k_{4}) \rangle = \frac{P_{\ell }^{\delta \delta }(k_{3})}{k_{3}^{2}}\delta^{1D}(k_{3}-k_{4})\delta_{\ell \ell '}^{K}\delta_{mm'}^{K} $, the covariance becomes
\begin{equation}
\begin{aligned}
C_{\ell }^{\Delta\Delta}(k,k')&=\int \mathrm{d}k_{1}k_{1}^{2} \int \mathrm{d}k_{2}k_{2}^{2} \int \mathrm{d}k_{3}^{2} Z_{\ell }(k,k_{1})Z_{\ell }(k',k_{2}) M_{\ell }(k_{1},k_{3})M_{\ell }(k_{2},k_{3}) P_{\ell }^{\delta\delta}(k_{3}).
\label{covarianceneutralhydrogenpowerspectrum}
\end{aligned}
\end{equation}

\subsection{The lensing potential}
As it was discussed previously, scalar fields on the sky that are associated with large-scale structures can be interpreted as line-of-sight integrated functions of the gravitational potential $\Phi$ with given weight. For the weak lensing scenario, we take a lensing potential $\phi$ for a given source at a 3d position in comoving distance $\textbf{r}=(\chi,\theta,\phi)$ to the gravitational potential by
\begin{equation}
\phi(\textbf{r}) = 
\phi(\chi,\theta,\phi) = 
\frac{2}{c^{2}}\int_{0}^{\chi}\mathrm{d}\chi'\left[ \frac{\chi-\chi'}{\chi\chi'} \right] \Phi(\chi',\theta,\phi).
\label{lensingpotential1}
\end{equation}
 where $F_{k}$ is determined by the curvature, and defined as $F_{K}(\chi) = \chi  $ for $K=0$. The density contrast $\delta(\textbf{r})$ is directly related to the gravitational potential $\Phi$ via Poisson equation
\begin{equation}
\nabla^{2}\Phi(\chi,\boldsymbol{\hat{n}})=\frac{3\Omega_{m}H_{0}^{2}}{2a(\chi)}\delta(\chi,\boldsymbol{\hat{n}})
\end{equation}
 This permits us to connect directly the statistics of the weak lensing observable to the underlying statistics of the mass distribution, and hence the cosmological parameters. In our convention $\chi=\chi(t)$ is the comoving distance to the emission source whose observed light was emitted at a given instance of time $t$ or redshift $z$. By using the 3-dimensional expansion, expressed in equations \ref{transform} and \ref{inversetransform}, the lensing potential can written, substituting equation \ref{lensingpotential1} into equation \ref{transform}, as
\begin{equation}
\phi_{\ell m}(k) = 
\sqrt{\frac{2}{\pi}} \frac{2}{c^{2}} \int \mathrm{d} \boldsymbol{r} \int \mathrm{d}\chi' \left[\frac{\chi-\chi'}{\chi\chi'}\right]j_{\ell }(k\chi) Y_{\ell m}^{*}(\theta,\phi)\Phi(\textbf{r'})
\label{philesing}
\end{equation}
 In the harmonic expansion, the Poisson equation can be expressed as $\Phi_{\ell m}(k; \chi)=\frac{A\delta_{\ell m}(k;\chi)}{a(\chi)k^{2}},$. Thus,
\begin{equation}
P_{\ell }^{\Phi \Phi}(k) = 
\left(\frac{3\Omega_{m}}{2\chi_{H}^{2}a k^{2}}\right)^{2}P_{\ell }^{\delta\delta}(k),
\end{equation}
 where we have introduced the Hubble distance $\chi_{H}=c/H_{0}$ and $A=-3\Omega_{m}/2\chi_{H}^{2} $. Here, $\Phi_{\ell m}(k) $ and $\delta_{\ell m}(k)$ are the spherical harmonic decomposition of $\Phi(\textbf{r})$ and $\delta(\textbf{r})$, respectively . The term $\chi$ dependence denotes the time-dependence of the potentials, which translates to a dependence on $\chi$, as $\chi$ corresponds to conformal lookback-time.
 
Equation \ref{philesing} can be simplified as 
\begin{equation}
\phi_{\ell m}(k) = \frac{A}{c^{2}}\int_{0}^{\infty}\mathrm{d}k'k'^{2}\eta_{\ell }(k,k')\frac{\delta_{\ell m}(k')}{k'^{2}}
\label{mainlensingequation}
\end{equation}
 where 
\begin{equation}
\begin{aligned}
\eta_{\ell }(k,k')&=\frac{4}{\pi}\int_{0}^{\infty}\mathrm{d}\chi \chi^{2}j_{\ell }(k\chi)\int_{0}^{\chi} \mathrm{d}\chi'\left[ \frac{\chi-\chi'}{\chi\chi'} \right] j_{\ell }(k'\chi')\frac{D[a(\chi')]}{a(\chi')}.
\end{aligned}
\end{equation}
 Furthermore, we can compute the 3d power spectrum of the lensing potential by using equation \ref{mainlensingequation}. This is given by
\begin{equation}
\begin{aligned}
C_{\ell }^{\phi\phi}(k_{1},k_{2})& = \frac{A^{2}}{c^{4}}\int_{0}^{\infty} \frac{\mathrm{d}k}{k^{2}} \eta_{\ell}(k_{1},k)\eta_{\ell }(k_{2},k) P^{\delta \delta}_{\ell }(k)= \int_{0}^{\infty}\mathrm{d}k k^{2}\eta_{\ell}(k_{1},k)\eta_{\ell }(k_{2},k)P^{\Phi \Phi}_{\ell }(k).
\label{lensingpowerspectrum}
\end{aligned}
\end{equation}
 In a complete form, we have 
\begin{equation}
\begin{aligned}
C_{\ell }^{\phi \phi}(k_{1},k_{2})& = \frac{16A^{2}}{c^{4}\pi^{2}} \int_{0}^{\infty}\mathrm{d}\chi_{a}\chi_{a}^{2}j_{\ell }(k_{1}\chi_{a}) \int_{0}^{\chi}\mathrm{d}\chi'_{a}\left[ \frac{\chi-\chi'_{a}}{\chi\chi'_{a}} \right] \int_{0}^{\infty}\mathrm{d}\chi_{b}\chi_{b}^{2}j_{\ell }(k_{2}\chi_{b}) \int_{0}^{\chi}\mathrm{d}\chi'_{b}\left[ \frac{\chi-\chi'_{b}}{\chi\chi'_{b}} \right]\\ & \times  \int_{0}^{\infty}  \frac{\mathrm{d}k}{k^{2}} P^{\delta \delta}_{\ell }(k,\chi'_{a},\chi'_{b}) j_{\ell }(k\chi'_{a})j_{\ell }(k\chi'_{b})\frac{D[a(\chi'_{a})] D[a(\chi'_{b})]}{a(\chi'_{a})a(\chi'_{b})}.
\label{covariancelensingpotential}
\end{aligned}     
\end{equation} 
 The above equation is called the covariance of the lensing  potential in 3 dimensions where shows some similarities to the expression obtained by \citet{PhysRevD.72.023516}. It must be pointed out that this covariance is a fully 3-dimensional quantity and is not longer homogeneous due to the radial dependence of the field.

From the previous derivations, we can already give an expression for the covariance of the 21 cm lensing in terms of the linear matter power spectrum $P(k)$, including the integrals $Z_{\ell}$, $M_{\ell}$ and $\eta_{\ell}$. The covariance of the deflection angle can be expressed as
\begin{equation}
\begin{aligned}
C_{\ell}^{\alpha\alpha}(k,k')=\left(\frac{3\Omega_{m}}{2\chi_{H}^{2}a }\right)^{2}\int \mathrm{d}k_{1}k_{1}^{2} \mathrm{d}k_{2}k_{2}^{2}\mathrm{d}k_{3}k_{3}^{2} \mathrm{d}k_{4} k_{4}^{2}\mathrm{d}k_{5}Z_{\ell}(k,k_{1})M_{\ell}(k_{1},k_{3})\eta_{\ell}(k_{3}, k_{5})\frac{P_{\ell}^{\delta\delta}(k_{5})}{k_{5}^{2}}\eta_{\ell}(k_{4}, k_{5})M_{\ell}(k_{2},k_{4})Z_{\ell}(k',k_{2}),
\end{aligned}\label{covariancedeflectionangle}
\end{equation} 
 where the integrals are defined by
\begin{equation}
    Z_{\ell}(k,k')=\frac{2}{\pi}\int \mathrm{d}s s^{2}\int \mathrm{d}\chi p(\chi|s)j_{\ell}(k'\chi)j_{\ell}(ks),\quad  M_{\ell}(k,k')=\frac{2}{\pi}\int \mathrm{d}\chi \chi^{2}j_{\ell}(k'\chi)j_{\ell}(k\chi)\tilde{W}(\chi)
\end{equation} 
and
\begin{equation}
\eta_{\ell }(k,k') =\frac{4}{\pi} \int_{0}^{\infty}\mathrm{d}\chi \chi^{2}j_{\ell}(k\chi)\int_{0}^{\chi}\mathrm{d}\chi' \left[ \frac{\chi-\chi'}{\chi\chi'} \right] j_{\ell}(k'\chi')\frac{D[a(\chi')]}{a(\chi')},
\end{equation} 
where $\eta_{\ell}(k,k')$ is mode coupling induced by lensing, $Z_{\ell}(k,k')$ encodes the redshift uncertainty by the probability distribution $p(\chi|s)$ and $M_{\ell}(k,k')$ is the radiation source distribution in distance encoded in the visibility function $\tilde{W}(\chi)$. One should notice that due to the strong oscillation of the spherical Bessel functions, the covariance will tend to fall off rapidly away from the diagonal $k = k'$.

\section{weak lensing analysis of 21-cm radiation}\label{analytical expressions}
The harmonic decomposition jointly uses ordinary spherical harmonics and spherical Bessel functions.
In the analysis of the 21-cm background, it is natural to apply a similar formalism as the one used in CMB lensing. If the lensing is weak in the sense that typical deflections are small compared to structure in the source field, the observed overdensity fluctuations can be expressed as a Taylor expansion of the unlensed temperature, where the effect of lensing changes angular components so radial components are unchanged
\begin{equation}
\begin{aligned}
\tilde{\Delta}(\chi,\boldsymbol{\hat{n}})&=\Delta(\chi, \boldsymbol{\hat{n}}+\nabla_{\boldsymbol{\hat{n}}} \phi(\chi,\boldsymbol{\hat{n}})) \approx \Delta(\chi,\boldsymbol{\hat{n}}) + \nabla_{\boldsymbol{\hat{n}}}\phi(\chi, \boldsymbol{\hat{n}})\nabla_{\boldsymbol{\hat{n}}}\Delta(\chi,\boldsymbol{\hat{n}})+O(\phi^{2}),
\end{aligned}\label{expansion}
\end{equation}
where the left-hand side represents the observed neutral hydrogen overdensity field. On the right-hand side, the first and second terms of the above expansion represent the unlensed neutral hydrogen density and the change of neutral hydrogen density field due to lensing at first order, respectively. The Taylor expansion used previously is valid in the CMB case due to the smallness of the temperature gradients on medium scales and Silk damping on smaller scales. This expansion is also valid in the 21-cm case, where the temperature gradients could be large, but the deflections are very small. By expanding the first term and the product of two scalar fields with their associated gradients  . In addition to expanding on the surface of the celestial sphere, we will also consider the expansion in the radial direction using spherical Bessel functions. So each side of equation~\ref{expansion} can be expressed in terms of spherical harmonics and Bessel functions:
\begin{equation}
\begin{aligned}
&\sum_{\ell=0}^{\infty}\sum_{m=-\ell}^{\ell}\tilde{\Delta}_{\ell m}(\chi)Y_{\ell m}(\boldsymbol{\hat{n}}) =\sum_{\ell ',m'}\Delta(\chi)_{\ell 'm'}Y_{\ell 'm'}(\boldsymbol{\hat{n}}) +  \sum_{\ell'=0}^{\infty}\sum_{m'=-\ell'}^{\ell'}\phi_{\ell'm'}(\chi)  \nabla_{\boldsymbol{\hat{n}}} Y_{\ell'm'}(\boldsymbol{\hat{n}}) \sum_{\ell''=0}^{\infty}\sum_{m''=-\ell''}^{\ell''} \Delta_{\ell''m''}(\chi)\nabla_{\boldsymbol{\hat{n}}} Y_{\ell''m''}(\boldsymbol{\hat{n}}) 
\end{aligned} 
\label{secondorder}
\end{equation}
with 
\begin{equation}
\Delta_{\ell''m''}(\chi)=\sqrt{\frac{2}{\pi}}\int_{0}^{\infty}k^2 \mathrm{d}k \Delta_{\ell''m''}(k)j_{\ell''}(k\chi),\quad  \phi_{\ell'm'}(\chi)=\sqrt{\frac{2}{\pi}}\int_{0}^{\infty}k^2 \mathrm{d}k\phi_{\ell'm'}(k)j_{\ell'}(k\chi)
\label{radial1}
\end{equation}
 Replacing the dummy indices in the sum on the left-hand side from $\ell m\rightarrow \ell'm'$, multiplying  $Y_{\ell m}^{*}$ on both sides and integrating both sides over all angles we obtain:
\begin{equation}
\begin{aligned}
\sum_{\ell '=0}^{\infty}\sum_{m'=-\ell '}^{\ell '}\tilde{\Delta}_{\ell 'm'}(\chi)\int \mathrm{d}\boldsymbol{\hat{n}}Y_{\ell 'm'}(\boldsymbol{\hat{n}})Y^{*}_{\ell m}(\boldsymbol{\hat{n}})&= \sum_{\ell ',m'}\Delta_{\ell 'm'}(\chi)\int \mathrm{d}\boldsymbol{\hat{n}} Y_{\ell 'm'}(\boldsymbol{\hat{n}}) Y^{*}_{\ell m}(\boldsymbol{\hat{n}}) \\ &+  \sum_{\ell'=0}^{\infty}\sum_{m'=-\ell '}^{\ell '}\phi_{\ell'm'}(\chi)   \sum_{\ell''=0}^{\infty}\sum_{m''=-\ell''}^{\ell''} \Delta_{\ell''m''}(\chi) \int \mathrm{d}\boldsymbol{\hat{n}}\nabla_{\boldsymbol{\hat{n}}} Y_{\ell'm'}(\boldsymbol{\hat{n}})\nabla_{\boldsymbol{\hat{n}}} Y_{\ell''m''}(\boldsymbol{\hat{n}}) Y^{*}_{\ell m}(\boldsymbol{\hat{n}})
\end{aligned}
\end{equation} 
 Applying the orthonormality relation of the spherical harmonics leads to
\begin{equation}
\begin{aligned}
\tilde{\Delta}_{\ell m}(\chi)& =\Delta_{\ell m}(\chi) +  \sum_{\ell '=0}^{\infty}\sum_{m'=-\ell '}^{\ell '}\phi_{\ell 'm'}(\chi)   \sum_{\ell ''=0}^{\infty}\sum_{m''=-\ell ''}^{\ell ''} \Delta_{\ell ''m''}(\chi)\int \mathrm{d}\boldsymbol{\hat{n}}\nabla_{\boldsymbol{\hat{n}}} Y_{\ell 'm'}(\boldsymbol{\hat{n}})\nabla_{\boldsymbol{\hat{n}}} Y_{\ell ''m''}(\boldsymbol{\hat{n}}) Y^{*}_{\ell m}(\boldsymbol{\hat{n}}).
\end{aligned}
\label{mainexpasionSFB}
\end{equation} 
 The last integral can be solved analytically. One can use the properties of the spherical harmonics and rewrite the last integral as
\begin{equation}
\begin{aligned}
     \int \mathrm{d}\boldsymbol{\hat{n}}
     \nabla_{\boldsymbol{\hat{n}}} Y_{\ell 'm'}(\boldsymbol{\hat{n}})\nabla_{\boldsymbol{\hat{n}}} Y_{\ell ''m''}(\boldsymbol{\hat{n}}) Y^{*}_{\ell m}(\boldsymbol{\hat{n}})& = \frac{1}{2} \int \mathrm{d}\boldsymbol{\hat{n}} [ Y_{\ell 'm'}(\boldsymbol{\hat{n}}) Y_{\ell ''m''}(\boldsymbol{\hat{n}})\nabla^{2}Y^{*}_{\ell m}(\boldsymbol{\hat{n}})\\ &-Y_{\ell 'm'}(\boldsymbol{\hat{n}})Y^{*}_{\ell m}(\boldsymbol{\hat{n}})\nabla^{2}Y_{\ell ''m''}(\boldsymbol{\hat{n}}) -Y_{\ell''m''}(\boldsymbol{\hat{n}})Y^{*}_{\ell m}(\boldsymbol{\hat{n}})\nabla^{2}Y_{\ell'm'}(\boldsymbol{\hat{n}}) ]
\end{aligned}
\end{equation}
 and use the identity $ \nabla^{2}Y_{\ell m}=-\ell(\ell+1)Y_{\ell m} $ to find
\begin{equation}
\begin{aligned}
&\int \mathrm{d}\boldsymbol{\hat{n}}
     \nabla_{\boldsymbol{\hat{n}}} Y_{\ell'm'}(\boldsymbol{\hat{n}})\nabla_{\boldsymbol{\hat{n}}} Y_{\ell''m''}(\boldsymbol{\hat{n}}) Y^{*}_{\ell m}(\boldsymbol{\hat{n}})= \frac{1}{2}[-\ell(\ell+1)+\ell'(\ell'+1)+\ell''(\ell''+1)] \int \mathrm{d}\boldsymbol{\hat{n}}
     Y^{*}_{\ell m}(\boldsymbol{\hat{n}})
     Y_{\ell'm'}(\boldsymbol{\hat{n}}) Y_{\ell''m''}(\boldsymbol{\hat{n}}).
\end{aligned}    
\end{equation} 
 Furthermore, one can use the Gaunt integral which has the following solution:
\begin{equation}
\begin{aligned}
&\int  \mathrm{d}\boldsymbol{\hat{n}} Y_{\ell m}^{*}(\boldsymbol{\hat{n}})  Y_{\ell'm'}(\boldsymbol{\hat{n}}) Y_{\ell''m''}(\boldsymbol{\hat{n}})= (-1)^{m}\sqrt{\frac{(2\ell+1)(2\ell'+1)(2\ell''+1)}{4\pi }} \begin{pmatrix} 
\ell & \ell' & \ell'' \\
0 & 0  & 0
\end{pmatrix}
\begin{pmatrix} 
\ell & \ell'& \ell'' \\
-m & m'& m'' 
\end{pmatrix},
\end{aligned}
\label{expression1}
\end{equation}
 leading to the following expression
\begin{equation}
\begin{aligned}
\int  \mathrm{d}\boldsymbol{\hat{n}} Y_{\ell m}^{*}(\boldsymbol{\hat{n}}) \nabla Y_{\ell 'm'}(\boldsymbol{\hat{n}}) \nabla Y_{\ell ''m''}(\boldsymbol{\hat{n}})&=\frac{1}{2}(-1)^{m} \left[\ell ''(\ell''+1)+\ell'(\ell'+1)-\ell(\ell+1) \right] \\ & \times \sqrt{\frac{(2\ell+1)(2\ell'+1)(2\ell''+1)}{4\pi }} \begin{pmatrix} 
\ell & \ell' & \ell'' \\
0 & 0  & 0
\end{pmatrix}
\begin{pmatrix} 
\ell & \ell'& \ell'' \\
-m & m'& m'' 
\end{pmatrix},
\end{aligned}
\label{expression2}
\end{equation} 
 where the second row of the first matrix indicates the spin-numbers of spherical harmonics which, in this case, are all zero. Substituing \ref{expression1} and \ref{expression2} into \ref{mainexpasionSFB}, equation \ref{expansion} turns into
\begin{equation}
\begin{aligned}
\tilde{\Delta}_{\ell m}(\chi)&=\Delta_{\ell m}(\chi) + \sum_{\ell'=0}^{\infty}\sum_{m'=-\ell'}^{\ell'}\sum_{\ell''=0}^{\infty}\sum_{m''=-\ell''}^{\ell''}\phi_{\ell'm'}(\chi) (-1)^{m}\begin{pmatrix} 
\ell & \ell'& \ell'' \\
-m & m'& m'' 
\end{pmatrix}
D_{\ell \ell'\ell''}\Delta_{\ell''m''}(\chi),
\end{aligned} 
\label{secondorder2}
\end{equation} 
 with the following  abbreviation
\begin{equation}
\begin{aligned}
  D_{\ell \ell'\ell''}=&\frac{1}{2}(-1)^{m} \left[\ell''(\ell''+1)+\ell'(\ell'+1)-\ell(\ell+1) \right] \sqrt{\frac{(2\ell+1)(2\ell'+1)(2\ell''+1)}{4\pi }} \begin{pmatrix} 
\ell & \ell' & \ell'' \\
0 & 0  & 0
\end{pmatrix}.
\end{aligned}
\label{abreviacion}
\end{equation}
 Finally combining expressions equation \ref{radial1} and the above definition, we find that the change to the neutral hydrogen overdensity fluctuation moments $\delta\tilde{\Delta}_{\ell m}(k)=\tilde{\Delta}_{\ell m}(k)-\Delta_{\ell m}(k) $ given by:
\begin{equation}
\begin{aligned}
\delta\tilde{\Delta}_{\ell m}(k)&= \left(\frac{2}{\pi}\right)^{3/2}\int \chi^{2}\mathrm{d}\chi j_{\ell}(k\chi) \sum_{\ell'm'}\sum_{\ell''m''}(-1)^{m} \begin{pmatrix}
\ell & \ell'& \ell'' \\
-m & m'& m'' 
\end{pmatrix}
 D_{\ell \ell'\ell''}\int_{0}^{\infty}k_{1}^{2}\mathrm{d}k_{1}j_{\ell'}(k_{1}\chi)\phi_{\ell'm'}(k_{1})\int_{0}^{\infty}k_{2}^{2}\mathrm{d}k_{2}j_{\ell''}(k_{2}\chi)\Delta_{\ell''m''}(k_{2}). 
\end{aligned}
\label{fullSFBdecomposition}
\end{equation} 
 Equation \ref{fullSFBdecomposition} is the  spherical Fourier-Bessel decomposition of the lensed field  $\tilde{\Delta}(\chi,\boldsymbol{\hat{n}})$. From the previous result, the covariance $C_{\ell}(k,k')$ of the neutral hydrogen overdensity field can be derived . Symbolically, in the calculation of covariance, terms such as $\langle \nabla \Delta \nabla \phi \rangle$, $\langle \Delta \nabla \phi \rangle$ are present but directly neglected since weak lensing and neutral hydrogen overdensity fields are largely separated and weakly correlated: A counter-example would be the integrated Sachs-Wolfe effect generated in the potentials responsible for gravitational lensing.

\subsection{Covariance of 21-cm brightness temperature with observable effects}
With the analysis of deep and large sections of the sky, the flat sky assumptions may not be regarded as the best way to manage incoming data. The 3d SFB decomposition arises as a natural basis for the source radiation on the spherical sky. In the context of the 21-cm radiation background, the signal is regarded continuous as a result of the unresolved mapping of the 21-cm brightness temperature. The radial component of the brightness temperature field $T_{b}(\chi_{\nu})$ is labeled by a given frequency since every redshift corresponds to a certain frequency so different radial distances from the observer. Since the fluctuations in temperature are the observable of interest here, we assume that the brightness temperature observed at a frequency $\nu$ in a certain radial survey along a direction $\boldsymbol{\hat{n}}$ is given by
\begin{equation}
    T_{b}(\boldsymbol{r})=T_{b}(\chi_{\nu},\boldsymbol{\hat{n}})=\overline{T}(\chi_{\nu})\Big(\Delta_{HI}(\chi_{\nu},\boldsymbol{\hat{n}})+1\Big),
    \label{brightness temperature field}
\end{equation} 
  where $\overline{T}(\chi_{\nu})$ is the average brightness temperature over the sky and $\chi_{\nu}$ is the comoving radial distance $ \chi(\nu)=\chi_{\nu}=\frac{c}{H_{0}}\int_{0}^{z(\nu)}\frac{\mathrm{d}z'}{E(z')}$ where $c$ is the speed of light, $H_{0}$ is the present-day Hubble parameter, with $1+z=\frac{\nu_\mathrm{rest}}{\nu} $ and $E(z)=[\Omega_{m}(1+z)^{3}+\Omega_{\Lambda}]^{1/2} $ , where $\nu_\mathrm{rest}$ is the rest frequency of the spectral line, $z$ is the redshift and $\Omega_{m}$ is the normalised matter density. At the beginning of last section, in order to avoid the information loss from the discrete binning of the observations, we decomposed the density as a 3-dimensional field into spherical harmonics and spherical Bessel functions. We can express the neutral hydrogen overdensity field $\Delta_{HI}(\boldsymbol{r})= \tilde{W}_{\Delta}(s_{\nu},\boldsymbol{\hat{n}})\delta(\chi_{\nu},\boldsymbol{\hat{n}})$, where $\tilde{W}_{\Delta}(s_{\nu},\boldsymbol{\hat{n}})$ is the visibility function with an estimated radial position with frequency $\nu$ and angular part, and $\delta(\boldsymbol{r})$ is the matter density fluctuation. Furthermore, we can include the conditional probability $p(\chi_{\nu}|s_{\nu})$ which relates the true and the estimated comoving radial distances. Thus, the term related to overdensity field equation in \ref{brightness temperature field} is rewritten as
\begin{equation}
\begin{aligned}
    T_{\ell m}(k)=&\sqrt{\frac{2}{\pi}}\int \mathrm{d}s s_{\nu}^{2}\int \mathrm{d}\chi p(\chi_{\nu}|s_{\nu}) \tilde{W}_{\Delta}(s_{\nu},\boldsymbol{\hat{n}})\overline{T}(\chi_{\nu}) b_{HI}D_{+}\delta(\chi_{\nu},\boldsymbol{\hat{n}}) j_{\ell }(ks_{\nu})  \int \mathrm{d}\boldsymbol{\hat{n}}Y_{\ell m}^{*}(\boldsymbol{\hat{n}}).
\end{aligned}
\label{temperaturedecomposition}
\end{equation}
As we stated in the previous section, with uncertain redshift estimates, $W_{\Delta}$ is smoothed and it can be approximated by the smoothed number density, $\tilde{W}_{\Delta}^{s}(\boldsymbol{r})$, at $\boldsymbol{r}$. Furthermore we denoted the transform of $\tilde{W}_{\Delta}(s_{\nu},\boldsymbol{\hat{n}})\delta(\chi_{\nu},\boldsymbol{\hat{n}})$ by $h_{\ell m}(k)$. For deep surveys, we separate the neutral hydrogen fraction visibility into a radial part and an angular selection as $\tilde{W}_{\Delta}^{s}(\chi_{\nu},\boldsymbol{\hat{n}})=\tilde{W}(\chi_{\nu})M(\boldsymbol{\hat{n}})$. So we can rewrite \ref{temperaturedecomposition} as
\begin{equation}
    \begin{aligned}
    T_{\ell m}(k)&=\frac{2}{\pi} \int \mathrm{d}s s_{\nu}^{2}\int \mathrm{d}\chi \overline{T}(\chi_{\nu}) p(\chi_{\nu}|s_{\nu})\sum_{\ell=0}^{\infty}\sum_{m'=-\ell'}^{\ell'} \int \mathrm{d}k'k'^{2}h_{\ell'm'}(k')j_{\ell'}(k'\chi_{\nu})j_{\ell}(ks_{\nu}) \int \mathrm{d}\boldsymbol{\hat{n}}Y_{\ell'm'}(\boldsymbol{\hat{n}}) Y_{\ell m}^{*}(\boldsymbol{\hat{n}})
    \\&=\int \mathrm{d}k'k'^{2}Z_{\ell}(k,k')h_{\ell m}(k'),
    \end{aligned}
\end{equation}\
 with $ Z_{\ell}(k,k')=\frac{2}{\pi} \int \mathrm{d}s s_{\nu}^{2}\int \mathrm{d}\chi p(\chi_{\nu}|s_{\nu})j_{\ell'}(k'\chi_{\nu})j_{\ell}(ks_{\nu})$. Calculating $h_{\ell m}(k')$ by direct substitution  of the expansion of $\delta(\chi_{\nu},\boldsymbol{\hat{n}})$, leading to
\begin{equation}
    h_{\ell m}(k')=\int \mathrm{d}k''k''^{2} M_{\ell}(k',k'')\delta_{\ell m}(k'')
\end{equation}
 where $ M_{\ell}(k',k'')=\frac{2}{\pi}\int \mathrm{d}\chi \chi_{\nu}^{2}\overline{T}(\chi_{\nu})b_{HI}D_{+}j_{\ell}(k''\chi_{\nu})j_{\ell}(\chi_{\nu}k')\tilde{W}(\chi_{\nu})$. Therefore, we can express the temperature coefficients as
\begin{equation}
    T_{\ell m}(k)=\int \mathrm{d}k' k'^{2}\int \mathrm{d}k'' k''^{2}Z_{\ell}(k,k')M_{\ell}(k',k'')\delta_{\ell m}(k'')
\end{equation}
 where $\delta_{\ell m}(k'')$ is the angular coefficient of the matter density fluctuations evaluated at certain wavenumber $k''$. The 3-dimensional power spectrum  is then obtained as the expectation value of two spherical Fourier-Bessel coefficients, 
\begin{equation}
\begin{aligned}
    C_{\ell}^{TT}(k,k')& =\Big\langle T_{\ell m}(k) T^{*}_{\ell' m'}(k') \Big\rangle =\int \mathrm{d}k_{1} k_{1}^{2}\mathrm{d}k_{2}k_{2}^{2}\mathrm{d}k_{3}k_{3}^{2}Z_{\ell}(k,k_{1})M_{\ell}(k_{1},k_{3})P(k_{3}) M_{\ell}(k_{3},k_{2})Z_{\ell}(k_{2},k').
    \label{temperaturecovarianceZM}
\end{aligned}
\end{equation} 
 Computing this power spectrum is numerically difficult due to the rapid oscillating spherical Bessel functions being integrated when computing the integrals $M_{\ell}$ and $Z_{\ell}$. However, this computation is achieved by applying a numerical method called Levin's integration exposed in \citet{LEVIN199695}. Equation \ref{temperaturecovarianceZM} paves the way to derive the lensed covariance of the brightness temperature by using equations \ref{fullSFBdecomposition}, \ref{abreviacion} and \ref{temperaturecovarianceZM}. After some algebra, the lensed covariance of 21-cm brightness temperature is expressed by
\begin{equation}
\begin{aligned}
C_{\ell}^{\tilde{T}\tilde{T}}(k,k')&=C_{\ell}^{TT}(k,k') + \left( \frac{2}{\pi}\right)^{3} \frac{1}{2\ell+1}\int_{0}^{\infty}\chi^{2}\mathrm{d}\chi j_{\ell }(k\chi)\int_{0}^{\infty}\chi'^{2}\mathrm{d}\chi'j_{L}(k'\chi') \sum_{\ell'=0}^{\infty} \sum_{\ell''=|\ell-\ell'|}^{\ell+\ell'}[D_{\ell \ell' \ell''}]^{2}  \\ &  \times\int_{0}^{\infty} k_{1}^{2}k'^{2}_{1}\mathrm{d}k_{1}\mathrm{d}k'_{1}j_{\ell'}(k_{1}\chi)j_{\ell'}(k'_{1}\chi') C_{\ell'}^{\phi\phi}(k_{1},k'_{1})\int_{0}^{\infty} k_{2}^{2}k'^{2}_{2} \mathrm{d}k_{2}\mathrm{d}k'_{2}j_{\ell''}(k_{2}\chi)j_{\ell''}(k'_{2}\chi') C^{TT}_{\ell''}(k_{2},k'_{2}).
\end{aligned}    
\end{equation}

\subsection{The 3d angular 21-cm power spectrum}
In a similar fashion to CMB computations of the angular power spectrum, we can also provide a first expression for the angular power spectrum caused by the fluctuations of the 21-cm brightness temperature on the sky. We must point out that this first derivation does not include the observable effects. The detected brightness temperature fluctuation is given by
\begin{equation}
    T_{b}^\mathrm{dec}(\chi(z), \boldsymbol{\hat{n}})=\int\mathrm{d}z\: T_{b}(\chi(z),\boldsymbol{\hat{n}}).
\end{equation}
Here, the term $T_{b}^\mathrm{dec}$ describes the detected temperature field projected in a certain direction and thin frequency shell $\delta \nu $ on to the sky. The brightness temperature fluctuations depend on the underlying neutral hydrogen density field which is approximately the matter density and to first order with an identical linear growth, but possibly distributed as a biased tracer with a parameter $b_{HI}$. Furthermore, the peculiar velocities of HI are not taken into account here since we neglect clumped regions of HI and avoid cross-correlations. Since the frequency of the 21-cm observations at high redshifts is not the real one but an estimated, hence we set a visibility function $\tilde{W}_{\nu}(z)$ as function of redshift and its corresponding frequency. Hence, we have
\begin{equation}
\begin{aligned}
T_{b}(\chi(z),\boldsymbol{\hat{n}})&= \overline{T}_{b}(z)\Big(\Delta_{HI}(\chi(z),\boldsymbol{\hat{n}})+1\Big)=\overline{T}_{b}(z)\tilde{W}_{\nu}(z)b_{HI}(z)D_{+}(z)\delta(\boldsymbol{r})+ \overline{T}_{b}(z).
\end{aligned}
\end{equation} 
Here, we just care about the fluctuating part or first term of the above expression so we apply the Fourier transformation of the density fluctuations, 
\begin{equation}
    \delta(\boldsymbol{r})=\int \frac{\mathrm{d}^{3}k}{(2\pi)^{3}}\tilde{\delta}(\boldsymbol{k})e^{i\boldsymbol{k}\cdot \boldsymbol{r}}
\end{equation}
 and using the Rayleigh-decomposition in the Fourier modes, $\exp(\mathrm{i}\boldsymbol{k}\cdot\boldsymbol{r}) = 4\pi \sum_{\ell m}\mathrm{i}^{\ell} j_{\ell}(k\chi) Y_{\ell m}(\boldsymbol{\hat{k}})Y_{\ell m}^{*}(\boldsymbol{\hat{n}}) $, we get
\begin{equation}
\begin{aligned}
       T^\mathrm{dec}_{b}(\chi_{\nu},\boldsymbol{\hat{n}}) =& 4\pi\sum_{\ell m}i^{l}\int\mathrm{d}z\: \tilde{W}_{\nu}(z)\overline{T}_{b}(z)b_{HI}(z)D_{+}(z)\int \frac{\mathrm{d}^{3}k}{(2\pi)^{3}}\tilde{\delta}(\boldsymbol{k}) j_{\ell}(k\chi(z)) Y_{\ell m}(\boldsymbol{\hat{k}})Y_{\ell m}^{*}(\boldsymbol{\hat{n}}).
\end{aligned}
\end{equation} 
 In the spherical Fourier-Bessel basis, angular fluctuations can be expressed by expanding the observed signal in spherical harmonics, $a_{\ell m}(\nu)=\int d\boldsymbol{\hat{n}}Y_{\ell m}(\boldsymbol{\hat{n}})T_{b}^{dec}(\chi_{\nu},\boldsymbol{\hat{n}}) $, we can use the closure relation for spherical harmonics to get
\begin{equation}
    \begin{aligned}
       a_{\ell m}(\nu)= & 4\pi\sum_{lm}i^{\ell}\int\mathrm{d}z\: \tilde{W}_{\nu}(z)\overline{T}_{b}(z)b_{HI}(z)D_{+}(z) \int \frac{\mathrm{d}^{3}k}{(2\pi)^{3}}\tilde{\delta}(\boldsymbol{k}) j_{\ell}(k\chi(z)) Y_{\ell m}(\boldsymbol{\hat{k}}).
    \end{aligned}
\end{equation} 
 Hence, the angular 21-cm power spectrum, $C_{l}$, is defined in terms of the ensemble average of two harmonic coefficients, $\langle a_{\ell m}(\nu) a^{*}_{\ell'm'}(\nu')\rangle = \delta_{\ell \ell'}^{K}\delta_{mm'}^{K}C_{\ell}(\nu, \nu')$, where $\delta^{K}$ denotes the Kronecker delta function and assuming that the temperature field is statistically homogeneous. Using the above equations and the matter power spectrum relation, $\langle \tilde{\delta}(\boldsymbol{k}) \tilde{\delta}^{*}(\boldsymbol{k'}) \rangle = (2\pi)^{3}\delta^{D}(\boldsymbol{k}-\boldsymbol{k'})P(k)$, we obtain an expression for the angular power spectrum given by 
\begin{equation}
    \begin{aligned}
    C_{\ell}(\nu, \nu')= &\frac{2}{\pi}\int \mathrm{d}z\: \tilde{W}_{\nu}(z)\overline{T}_{b}(z)b_{HI}(z)D_{+}(z) \int \mathrm{d}z'\: \tilde{W}_{\nu'}(z')\overline{T}_{b}(z')b_{HI}(z')D_{+}(z') \int \mathrm{d}k k^{2} P(k) j_{\ell}(k\chi(z))j_{\ell}(k\chi(z')).
    \end{aligned}
    \label{3d21cmangularpowerspectrum}
\end{equation} 
Equation~\ref{3d21cmangularpowerspectrum} can also be seen as the multifrequency angular power spectrum of 21-cm  brigthness temperature fluctuations at two different frequencies $\nu$ and $\nu'$ and derived by following the line of reasoning in \cite{PhysRevD.95.063522}. It is noticed that as the value of $\nu'$ increases, the two spherical Bessel functions $j_{\ell}(k\chi(z)) $ and $j_{\ell}(k\chi(z'))$ oscillate out of phase. As a result the value of covariance $C_{\ell}$ is expected to fall as $\nu'$ increases. At very large multipoles, equation \ref{3d21cmangularpowerspectrum} becomes hard to compute due to the rapid oscillations of the spherical Bessel functions. One way to tackle this issue is to implement the Limber approximation, which is precise at large $\ell$ and easy to compute. The Limber approximation replaces the spherical Bessel function with a $\delta$-function,
\begin{equation}
j_{\ell}(k\chi) \rightarrow \sqrt{\frac{\pi}{2(\ell+1/2)}}\delta^{D}(\ell+1/2-k\chi),
\end{equation} 
 where the wavenumber $k$ is related to the radial comoving distance $\chi$ with the relation $k\chi=\ell+\frac{1}{2}$ \citep{PhysRevD.70.083536, PhysRevD.78.123506}. Therefore, the 3d-angular power spectrum becomes diagonal in frequency  and leads to
\begin{equation}
    C_{\ell}(\nu)=\int \mathrm{d}z \left
    (\frac{\tilde{W}_{\nu}(z)\overline{T}_{b}(z)b_{HI}(z)D_{+}(z)}{\chi(z)} \right)^{2}\frac{P\left[ \frac{\ell+\frac{1}{2}}{\chi(z)} \right]}{|\chi'(z)|}.
    \label{3dpowerspectrumlimberapprox}
\end{equation} 
 The 3d angular power spectrum given in equation \ref{3d21cmangularpowerspectrum} is different from that in reference \cite{FURLANETTO2006181}, where it involves a frequency response of the experiment.

\section{Quadratic lensing estimator for 21-cm fields}\label{quadratic estimator}

\subsection{Full-sky lensing reconstruction}
Because the 21-cm radiation background lensing generates a correlation between the temperature and its gradient, such couplings can be used to construct an estimator, quadratic in the observed temperature. Hence we derive the lensing quadratic estimator, applying a similar technique to those derived in \citet {Hu_2001, Hu_2002, PhysRevD.95.043508, PhysRevD.67.083002}. To perform this, we calculate the covariance of the lensed brightness temperature,
\begin{equation}
    \langle \tilde{T}_{LM}(k)\tilde{T}^{*}_{L'M'}(k') \rangle|_{\text{21-cm}},
\end{equation} 
 in the Fourier-Bessel harmonic space by using the SFB decomposition of the lensed field 
\begin{equation}
\begin{aligned}
\tilde{T}_{\ell m}(\chi)&=T_{\ell m}(\chi) + \sum_{\ell'=0}^{\infty}\sum_{m'=-\ell'}^{\ell'}\sum_{\ell''=0}^{\infty}\sum_{m''=-\ell''}^{\ell''}T_{\ell''m''}(\chi) (-1)^{m} \begin{pmatrix}
\ell & \ell'& \ell'' \\
-m & m'& m'' 
\end{pmatrix}
D_{\ell \ell'\ell''}\phi_{\ell'm'}(\chi).
\end{aligned} 
\label{secondorder2a}
\end{equation} 
 We should note two things: that the 3d correlation function is not homogeneous and the lensing potential is a function of $k$. The above equation can be reduced to
\begin{equation}
\begin{aligned}
\Big\langle \tilde{T}_{LM}(k)\tilde{T}^{*}_{L'M'}(k')\Big\rangle_{\text{21-cm}} &= C_{L}(k,k')\delta^{K}_{LL'}\delta^{K}_{MM'}+ \left(\frac{2}{\pi} \right)^{2}\sum_{\ell, m}(-1)^{m}\begin{pmatrix}
\ell & L & L' \\
-m &  M & M' 
\end{pmatrix} g_{\ell L L'}^{\phi,(TT)}(k,k')\phi_{\ell, m}(\kappa),
\label{fourier-bessel correlation2}
\end{aligned}    
\end{equation} 
 with the following abbreviation $g_{\ell L L'}^{\phi,(TT)}(k,k')=D_{L, \ell, L' }C_{L'}^{TT}(k,k')+  D_{L' \ell L } C_{L}^{TT}(k',k') $, where $D_{\ell \ell' \ell'' }$ and $C_{\ell}$ are given in equations \ref{abreviacion} and \ref{covarianceneutralhydrogenpowerspectrum}, respectively. To extract the off-diagonal terms and find the solutions for $\phi_{\ell m}(k)$, we multiply 
\begin{equation}
    (-1)^{m'}\begin{pmatrix} 
\ell' & L & L' \\
-m' & M & M' 
\end{pmatrix}g_{\ell' L L'}^{\phi,(TT)}(k,k')
\end{equation} 
 in both sides of Eq. \ref{fourier-bessel correlation2}. Then, summing up the equation over $M$ and $M'$, and using the Wigner-$3j$ symbols identities, we find \footnote{In the derivation of equation \ref{lensingestimator1}, we ignored the unlensed term $C_{L}$, coming from the first term in equation \ref{fourier-bessel correlation2}.}
\begin{equation}
\begin{aligned}
    \phi_{\ell m}(\kappa)&= \frac{2\ell+1}{g_{\ell L L'}^{\phi,(TT)}(k,k')}\sum_{M,M'}(-1)^{m}\begin{pmatrix} \ell & L & L' \\
-m & M & M' 
\end{pmatrix}\langle \tilde{T}_{L M}(k)\tilde{T}_{L'M'}^{*}(k')\rangle_{\text{21-cm}}.
\label{lensingestimator1}
\end{aligned}
\end{equation} 
The above expression cannot be regarded as our estimator of the lensing potential, because the equation has the ensemble average over the 21-cm radiation background brightness temperature alone, $\langle\ldots\rangle_{\text{21-cm}}$. However, it points out that, by summing the quadratic combination of lensed fields over the multipoles, it is possible to construct the estimator for the scalar lensing potential $\phi$. Based on our previous calculations, we can establish our estimator in the following forms
\begin{equation}
\begin{aligned}
  \hat{\phi}_{\ell m}(\kappa)&=\frac{2\ell+1}{g_{\ell L L'}^{\phi,(TT)}(k,k')} \sum_{M,M'}(-1)^{m}\begin{pmatrix} 
\ell & L & L' \\
-m & M & M' 
\end{pmatrix} \tilde{T}_{L M}(k)\tilde{T}_{L'M'}(k')=\phi_{\ell m}(\kappa) + n_{\ell,m,L,L'}^{\phi,(TT)}(\kappa),
\end{aligned}
\end{equation} 
 where $\tilde{T}$ represents the lensed temperature.We follow the general decomposition of the quadratic estimates
\begin{equation}
\begin{aligned}
 \langle \hat{\phi}_{\ell m}(\kappa) \hat{\phi}^{*}_{\ell' m'}(\kappa')\rangle &=\delta^{K}_{\ell\ell'}\delta^{K}_{mm'} \left(C_{\ell}^{\phi\phi}(\kappa,\kappa')+N_{\ell}^{0}(\kappa,\kappa')+ N_{\ell}^{1}(\kappa,\kappa') \right),
\end{aligned}
\end{equation} 
where the first term $N_{\ell}^{0}$ is related to the disconnected terms of the lensed 21-cm radiation background four-point correlation, whereas the higher order terms $N_{\ell}^{i}$ for $i\geq1$ are related to the connected terms. The estimator includes the contribution from the term $n_{\ell,m,L,L'}^{\phi,(TT)}(\kappa)$, which leads to the noisy reconstruction of the lensing potentials. To solve this puzzle, we propose to redefine the estimator $\phi$ by introducing a weight function $A_{\ell L L'}^{\phi}$, in order to reduce the contribution from $ n_{\ell,m,L,L'}^{\phi,(TT) }$. By summing up all the possible combinations of the angular components $L$ and $L'$, we express the estimator of the lensing potential, similar to that proposed in \cite{PhysRevD.95.043508}
\begin{equation}
\begin{aligned}
  \hat{\phi}^{TT}_{\ell m}(\kappa)= \sum_{L,L'}A_{\ell,L,L'}^{\phi,(TT)} \sum_{M,M'}(-1)^{m}\begin{pmatrix} \ell & L & L' \\
-m & M & M' 
\end{pmatrix} \tilde{T}
_{L M}(k)\tilde{T}_{L'M'}(k'),
\label{estimator}
\end{aligned} 
\end{equation} 
 where the form of the weight function must be determined so that the noise contribution is minimised. Following the recipe of \citet{Hu_2002}, we can express the estimator as
\begin{equation}
    \hat{\phi}^{TT}_{\ell m}(\kappa)= \sum_{L,L'}A_{\ell,L,L'}^{\phi,(TT)}\frac{g_{\ell L L'}^{\phi,(TT)}(k,k')}{2\ell+1}\phi_{\ell m}(\kappa)+n_{\ell m }^{\phi,(TT)}(\kappa),\quad \text{with} \quad n_{\ell m}^{\phi,(TT)}(\kappa)=\sum_{L,L'}A_{\ell,L,L'}^{\phi,(TT)}\frac{g_{\ell L L'}^{\phi,(TT)}(k_{1},k_{2})}{2\ell+1}n_{\ell,m,L,L'}^{\phi,(TT)}(\kappa)
    \label{guessestimator}
\end{equation} 
 Equation \ref{guessestimator} tells us that the estimator is an unbiased estimator if we establish the condition:
\begin{equation}
    \sum_{L,L'}A_{\ell,L,L'}^{\phi,(TT)}\frac{g_{\ell L L'}^{\phi,(TT)}(k_{1},k_{2})}{2\ell+1}=1, \quad \text{equivalent to}\quad \langle \hat{\phi}_{\ell m }(\kappa) \rangle_{\text{21-cm}}=\phi_{\ell m }(\kappa)
    \label{unbiasedcondition}
\end{equation}
At the same time, we would like to suppress the noise contribution, $n_{\ell m }^{\phi,(TT)}(\kappa)$, with the following condition $\frac{\delta \langle |n_{\ell m }^{\phi,(TT)}(\kappa)|^{2}\rangle}{\delta A_{\ell L L'}^{\phi,(TT)}}=0 $ so we calculate the form of the weight functions under the conditions established above, with the Lagrange-multiplier technique. The variance of the $n_{\ell m }^{\phi,(TT)}(\kappa)$ can be computed as follows. First, we re-express the noise variance using the variance of the estimator.  Let us set $n_{\ell m }^{\phi,(i)}(\kappa)=\hat{\phi}_{\ell m }^{(i)}(\kappa)-\phi_{\ell m }(\kappa)$ with $i, j=TT$. For now, let's drop the dependence of $\kappa$ for simplicity and we will take it back at the end of the calculation. Hence the noise variance is rewritten as
\begin{equation}
\begin{aligned}
    \Big\langle \left(n_{\ell m }^{\phi,(i)}\right)^{*} n_{\ell m }^{\phi,(j)} \Big\rangle =  \Big\langle \left(\hat{\phi}_{\ell m }^{(i)}\right)^{*}\hat{\phi}_{\ell m }^{(j)} \Big\rangle- \Big\langle \left(\phi_{\ell m }\right)^{*}\hat{\phi}_{\ell m }^{(j)}  \Big\rangle-\Big\langle  \left(\hat{\phi}_{\ell m }^{(i)}\right)^{*} \phi_{\ell m } \Big\rangle+ C_{\ell }^{\phi\phi}.
\end{aligned}
\end{equation} 
 We carefully need to note here that
\begin{equation}
\begin{aligned}
 \Big\langle\left(\phi_{\ell m }\right)^{*} \tilde{T}_{L M}(k)\tilde{T}_{L'M'}(k') \Big\rangle = \sum_{\ell'm'}\sum_{L''M''}&\Bigg[ (-1)^{M'} \begin{pmatrix} 
L' & \ell'& L'' \\
-M' & m' & M'' 
\end{pmatrix} D_{L' \ell'L''} \Big\langle \left(\phi_{\ell m}\right)^{*} T_{L''M''}(k'')\phi_{\ell'm'}T_{L M}(k) \Big\rangle \\& + (-1)^{M}\begin{pmatrix} 
L & \ell'& L'' \\
-M & m' & M'' 
\end{pmatrix} D_{L \ell'L''} \Big\langle \left(\phi_{\ell m }^{*}\right)T_{L''M''}(k'')\phi_{\ell'm'}T_{L'M'}(k') \Big\rangle \Bigg].
\end{aligned}
\end{equation} 
 Assuming that the lensing potential and 21-cm radiation background are Gaussian random fields, and the correlation between them are negligible, the above equation becomes
\begin{equation}
\begin{aligned}
  \Big\langle \left(\phi_{\ell m }\right)^{*} \tilde{T}_{LM}(k)\tilde{T}_{L'M'}(k')\ \Big\rangle &=(-1)^{m'} \begin{pmatrix} 
\ell & L & L' \\
-m & M & M'
\end{pmatrix} g_{\ell L L'}^{\phi,(TT)}(k,k')C_{\ell}^{\phi\phi}.
\end{aligned}
\end{equation} 
By using one of the Wigner-$3j$ symbols identities, this leads to the following expression
\begin{equation}
\begin{aligned}
    \Big\langle \left(\phi_{\ell m}\right)^{*}\hat{\phi}_{\ell m }^{i}\Big\rangle&= \sum_{L,L'}A_{\ell,L,L'}^{\phi,(TT)} \sum_{M,M'}\begin{pmatrix} 
\ell & L' & L' \\
-m & M & M'
\end{pmatrix}\begin{pmatrix} 
\ell & L & L' \\
-m & M & M'
\end{pmatrix} g_{\ell L L'}^{\phi,(TT)}(k,k')C_{\ell}^{\phi\phi}=C_{\ell}^{\phi\phi}.
\end{aligned}
\end{equation} 
Consequently, one obtains 
\begin{equation}
    \Big\langle \left(n_{\ell m }^{\phi,(i)}\right)^{*} n_{\ell m }^{\phi,(j)} \Big\rangle = \Big\langle \left(\hat{\phi}_{\ell m }^{(i)}\right)^{*}\hat{\phi}_{\ell m }^{(j)} \Big\rangle - C_{\ell }^{\phi\phi}.
\end{equation}
 Next we calculate the estimator covariance, $ \Big\langle \left(\hat{\phi}_{\ell m }^{(i)}\right)^{*}\hat{\phi}_{\ell m }^{(j)} \Big\rangle  $. From equation \ref{estimator}, the covariance is given by 
\begin{equation}
\begin{aligned}
     \Big\langle \left(\hat{\phi}_{\ell m }^{(i)}\right)^{*}\hat{\phi}_{\ell m }^{(j)} \Big\rangle& = \sum_{L_{1},L'_{1}} \sum_{L_{2},L'_{2}} \sum_{M_{1},M'_{1}} \sum_{M_{2},M'_{2}}\begin{pmatrix} 
\ell & L_{1}& L'_{1} \\
-m & M_{1} & M'_{1}
\end{pmatrix}\begin{pmatrix} 
\ell & L_{2}& L_{2}' \\
-m & M_{2} & M_{2}'
\end{pmatrix} \left( A_{\ell ,L_{1},L_{1}'}^{\phi,(i)} \right)^{*}A_{\ell,L_{2},L_{2}'}^{\phi,(j)} \Big\langle \tilde{T}^{*}_{L_{1}M_{1}}\tilde{T}^{*}_{L_{1}'M_{1}'} \tilde{T}_{L_{2}M_{2}}\tilde{T}_{L_{2}'M_{2}'} \Big\rangle.
\end{aligned}
\end{equation}
Here, we omit the dependence of the wavenumber $k$ in $\tilde{T}$ for simplicity. We have to compute without doubt the four-point correlation for random Gaussian fields of the possible observed 21-cm radiation background brightness temperature as \cite{Hu_2001,Hu_2002}:
\begin{equation}
\begin{aligned}
    \Big\langle \tilde{T}^{*}_{L_{1}M_{1}}\tilde{T}^{*}_{L_{1}'M_{1}'} \tilde{T}_{L_{2}M_{2}}\tilde{T}_{L_{2}'M_{2}'} \Big\rangle &= \tilde{C}_{L_{1}}^{TT} \tilde{C}_{L_{1'}}^{TT}\delta_{L_{1}L_{1}'}\delta_{L_{2}L_{2}'}\delta_{M_{1},-M_{1}'}\delta_{M_{2},-M_{2}'} +\tilde{C}_{L_{1}}^{TT}\tilde{C}_{L_{1'}}^{TT}\delta_{L_{1}L_{2}}\delta_{L_{1}'L_{2}'}\delta_{M_{1},M_{2}}\delta_{M_{1}',M_{2}'} \\&+ \tilde{C}_{L_{1}}^{TT}\tilde{C}_{L_{1'}}^{TT}\delta_{L_{1}L_{2}'}\delta_{L_{1}'L_{2}}\delta_{M_{1},M_{2}'}\delta_{M_{1}',M_{2}}
\end{aligned}
\end{equation} 
Using the summation of Wigner-$3j$ symbol, the first term gives $\delta_{\ell,0}$, and we neglect this term to consider $\ell>0$. Hence we obtain
\begin{equation}
\begin{aligned}
\Big\langle \left(\hat{\phi}_{\ell m }^{(i)}\right)^{*}\hat{\phi}_{\ell m}^{(j)} \Big\rangle = &\sum_{L_{1},L'_{1}} \sum_{L_{2},L'_{2}} \sum_{M_{1},M'_{1}} \sum_{M_{2},M'_{2}} \begin{pmatrix}
\ell  & L_{1}& L'_{1} \\
-m  & M_{1} & M'_{1}
\end{pmatrix}\begin{pmatrix} 
\ell & L_{2}& L_{2}' \\
-m & M_{2} & M_{2}'
\end{pmatrix} \left( A_{\ell,L_{1},L_{1}'}^{\phi,(i)} \right)^{*}A_{\ell,L_{2},L_{2}'}^{\phi,(j)}\\& \times\Bigg[\tilde{C}_{L_{1}}^{TT}\tilde{C}_{L_{1'}}^{TT}\delta_{L_{1}L_{2}}\delta_{L_{1}'L_{2}'}\delta_{M_{1},M_{2}}\delta_{M_{1}',M_{2}'}+ \tilde{C}_{L_{1}}^{TT}\tilde{C}_{L_{1'}}^{TT}\delta_{L_{1}L_{2}'}\delta_{L_{1}'L_{2}}\delta_{M_{1},M_{2}'}\delta_{M_{1}',M_{2}} \Bigg]\\& = \sum_{L_{1},L'_{1}} \sum_{M_{1},M'_{1}} \left( A_{\ell ,L_{1},L_{1}'}^{\phi,(i)} \right)^{*} \Bigg[ A_{\ell ,L_{1},L_{1}'}^{\phi,(j)}\tilde{C}_{L_{1}}^{TT}\tilde{C}_{L_{1'}}^{TT}\begin{pmatrix} 
\ell & L_{1}& L'_{1} \\
-m & M_{1} & M'_{1}
\end{pmatrix}\begin{pmatrix} 
\ell & L_{1}& L_{1}' \\
-m & M_{1} & M_{1}'
\end{pmatrix} \\& + A_{\ell,L_{1}',L_{1}}^{\phi,(j)}\tilde{C}_{L_{1}}^{TT}\tilde{C}_{L_{1'}}^{TT}\begin{pmatrix} 
\ell & L_{1}& L'_{1} \\
-m & M_{1} & M'_{1}
\end{pmatrix}\begin{pmatrix} 
\ell & L_{1}'& L_{1} \\
-m & M_{1}' & M_{1}
\end{pmatrix}  \Bigg].
\end{aligned}    
\end{equation} 
Again, using Wigner-$3j$ symbol identities, the above equation reduces to
\begin{equation}
\begin{aligned}
\Big\langle \left(\hat{\phi}_{\ell m }^{(i)}\right)^{*}\hat{\phi}_{\ell m}^{(j)} \Big\rangle & = \frac{1}{2\ell+1}\sum_{L_{1},L_{1}'}\left(A_{\ell,L_{1},L_{1}'}^{\phi,(i)} \right)^{*} \Bigg( A_{\ell,L_{1},L_{1}'}^{\phi,(j)}\tilde{C}_{L_{1}}^{TT}\tilde{C}_{L_{1'}}^{TT}+(-1)^{\ell+L_{1}+L_{1}'}A_{\ell,L_{1}',L_{1}}^{\phi,(j)} \tilde{C}_{L_{1}}^{TT}\tilde{C}_{L_{1'}}^{TT}  \Bigg)
\end{aligned}
\end{equation} 
 where the term $\tilde{C}_{\ell}^{TT}$ is the lensed covariance including the observable effects as redshift estimation and source distribution. After a harsh calculation, the condition becomes 
\begin{equation}
\begin{aligned}
    \frac{\delta}{\delta A_{\ell,L, L'}^{\phi,(TT)} }& \Bigg[ \frac{1}{2\ell+1}\sum_{L,L'} \left(A_{\ell,L,L'}^{\phi,(TT)} \right)^{*} \Bigg(  A_{\ell,L,L'}^{\phi,(TT)}\tilde{C}_{L}^{TT}\tilde{C}_{L'}^{TT}+(-1)^{\ell+L+L'}A_{\ell,L,L'}^{\phi,(TT)} \tilde{C}_{L}^{TT}\tilde{C}_{L'}^{TT}\Bigg) + \Lambda \left( \sum_{L,L'}A_{\ell,L,L'}^{\phi,(TT)}\frac{g_{\ell L L'}^{\phi,(TT)}(k,k')}{2\ell+1}-1    \right)\Bigg]=0
\end{aligned}
\end{equation} 
 where $\Lambda$ is the Lagrange multiplier that serves to minimise our quantity. The above equation leads to
\begin{equation}
\begin{aligned}
\left(A_{\ell,L,L'}^{\phi,(TT)} \right)^{*}\tilde{C}_{L}^{TT}\tilde{C}_{L'}^{TT}&+ \left(A_{\ell,L,L'}^{\phi,(TT)} \right)^{*}(-1)^{\ell+L+L'} \tilde{C}_{L}^{TT}\tilde{C}_{L'}^{TT}+\Lambda g_{\ell L L'}^{\phi,(TT)}(k,k')=0.
  \label{lagrange1}
\end{aligned}
\end{equation} 
 In the above, we exchange $L$ and $L'$, to arrive at
\begin{equation}
\begin{aligned}
\left(A_{\ell,L',L}^{\phi,(TT)} \right)^{*}\tilde{C}_{L'}^{TT}\tilde{C}_{L}^{TT}&+ \left(A_{\ell,L,L'}^{\phi,(TT)} \right)^{*}(-1)^{\ell+L+L'} \tilde{C}_{L}^{TT}\tilde{C}_{L'}^{TT}+\Lambda g_{\ell L' L}^{\phi,(TT)}(k,k')=0.
    \label{lagrange2}
\end{aligned}
\end{equation} 
 performing the product of the factors $\tilde{C}_{L'}^{TT}\tilde{C}_{L}^{TT}$ and $-(-1)^{\ell+L+L'}\tilde{C}_{L}^{TT}\tilde{C}_{L'}^{TT}$ with eq. \ref{lagrange1} and eq. \ref{lagrange2}, respectively, the sum of eqs.~\ref{lagrange1} and~\ref{lagrange2} gives 
\begin{equation}
    A_{\ell,L,L'}^{\phi,(TT)}+\Lambda f_{\ell L L'}^{\phi,(TT)}=0,
    \label{lagrangeequation}
\end{equation}
 with the abbreviation
\begin{equation}
\begin{aligned}
 f_{\ell L L'}^{\phi,(TT)}&=\frac{1}{\tilde{C}_{L}^{TT}\tilde{C}_{L'}^{TT}\tilde{C}_{L'}^{TT}\tilde{C}_{L}^{TT}-(\tilde{C}_{L}^{TT}\tilde{C}_{L'}^{TT})^{2}}\Bigg[ \left(g_{\ell,L,L'}^{\phi,(TT)}(k,k')\right)^{*}\tilde{C}_{L'}^{TT}\tilde{C}_{L}^{TT}-(-1)^{\ell+L+L'}\tilde{C}_{L}^{TT}\tilde{C}_{L'}^{TT}\left(g_{\ell,L',L}^{\phi,(TT)}(k,k')\right)^{*}\Bigg]
\end{aligned}
\end{equation} 
 Writing the abbreviation $(a^{\phi},b^{\phi})_{\ell}^{TT}$ as
\begin{equation}
    (a^{\phi},b^{\phi})_{\ell}^{TT}=\frac{1}{2\ell+1}\sum_{L,L'}a_{\ell,L,L'}^{\phi,(TT)}b_{\ell,L,L'}^{\phi,(TT)}
\end{equation}
and substituting equation~\ref{lagrangeequation} into equation \ref{unbiasedcondition}, we obtain the relation of the form
 \begin{equation}
     -\left(\Lambda\right)^{*}(f^{\phi},g^{\phi})_{\ell}^{TT}=1 \quad \rightarrow \quad \left(\Lambda\right)^{*} = -\frac{1}{(f^{\phi},g^{\phi})_{\ell}^{TT}}.
 \end{equation} 
Hence, from equation \ref{lagrangeequation}, we get the expression for the weight function: 
 \begin{equation}
     A_{\ell,L,L'}^{\phi,(TT)}=\frac{f_{\ell L L'}^{\phi,(TT)}}{(g^{\phi},f^{\phi})^{TT}_{\ell}}
 \end{equation} 
With the weight function already calculated, the noise variance, $N_{\ell}^{\phi,(TT)}$, becomes 
\begin{equation}
\begin{aligned}
N_{\ell}^{\phi,(TT)}&=\Big\langle |n_{\ell m }^{\phi,(TT)}|^{2}\Big\rangle= \frac{1}{2\ell+1}\frac{1}{(g^{\phi},f^{\phi})^{TT}_{\ell}}\sum_{L,L'}\left( f_{\ell L L'}^{\phi,(TT)} \right)^{*}\left( A_{\ell,L,L'}^{\phi,(TT)}\tilde{C}_{L}^{TT}\tilde{C}_{L'}^{TT}+ A_{\ell,L',L}^{\phi,(TT)}(-1)^{\ell+L+L'} \tilde{C}_{L}^{TT}\tilde{C}_{L'}^{TT}\right)=\frac{1}{(g^{\phi},f^{\phi})_{\ell}^{TT}}.
\end{aligned}     
\end{equation}
Therefore, the noise variance and weight function are given by
\begin{equation}
\begin{aligned}
 N_{\ell}^{\phi,(TT)}(\kappa)&= \Bigg[ \frac{1}{2\ell+1}\sum_{L,L'}g_{\ell L L'}^{\phi,TT}(k,k')\left(g_{\ell L L'}^{\phi,TT}(k,k') \right)^{*} \frac{1}{\tilde{C}_{L}^{TT}(k_{1},k_{1}')\tilde{C}_{L'}^{TT}(k_{2},k_{2}')} \Bigg]^{-1}, \quad  A_{\ell,L,L'}^{\phi,(TT)}=N_{\ell}^{\phi,(TT)}f_{\ell L L'}^{\phi,TT}.
\end{aligned}
\end{equation} 
It results that the full-sky lensing reconstruction from 21-cm radiation background includes the observable effects such as redshift estimation $Z_{\ell}$, source distribution $M_{\ell}$. The numerical computations of the quadratic lensing estimator, the noise variance and their comparisons with solutions of earlier works are not part of this analysis and it is left for a future work. However it is noticeable the similarities to those results derived in \citet{PhysRevD.67.083002, LEWIS20061, 10.1093/mnras/stz1781}.

\subsection{Flat-sky approximation}
Here, we will derive our quadratic estimator based on the flat-sky limit. In the flat sky, we expand a 3d field $f$ at 3d position $\boldsymbol{r}=(\chi,\boldsymbol{\hat{n}})$ on the sky into a combination of 2D Fourier modes and Bessel functions in the radial direction $\chi$
\begin{equation}
    f(\chi,\boldsymbol{\hat{n}})=\sqrt{\frac{2}{\pi}}\int_{0}^{\infty} k^{2}\mathrm{d}k\int_{0}^{\infty} \frac{\mathrm{d}^{2}\boldsymbol{\ell}}{(2\pi)^{2}}f(k,\boldsymbol{\ell})j_{\ell}(k\chi)e^{i\boldsymbol{\ell}\cdot\boldsymbol{\hat{n}}},
\end{equation}
\begin{equation}
f(k,\boldsymbol{\ell})=\sqrt{\frac{2}{\pi}}\int_{0}^{\infty}\chi^{2}\mathrm{d}\chi \int_{0}^{\infty}\mathrm{d}^{2}\boldsymbol{\hat{n}}f(\chi,\boldsymbol{\hat{n}})j_{\ell}(k\chi)e^{-i\boldsymbol{\ell}\cdot\boldsymbol{\hat{n}}},
\end{equation}
where such expansion keeps a relation with the 3d full sky expansion introduced previously. Therefore, it has the advantages as its full-sky counterpart. Here we would like to set a relation between the expansion coefficients $f(k,\boldsymbol{\ell})$ in the flat sky as defined above and $f_{\ell m}(k)$ in the full sky. The relation between the 3d flat-sky and the 3d full-sky coefficients are given by
\begin{equation}
    f(k,\boldsymbol{\ell})=\sqrt{\frac{2\pi}{\ell}}\sum_{m}i^{-m}f_{\ell m}(k)e^{im\varphi_{\ell}}.
    \label{3dflat2}
\end{equation}
\begin{equation}
    f_{\ell m}(k)=\sqrt{\frac{\ell}{2\pi}}i^{m}\int_{0}^{2\pi}\frac{\mathrm{d}\varphi_{\ell}}{2\pi}e^{-im\varphi_{\ell}}f(k,\boldsymbol{\ell}),
    \label{3dflat1}
\end{equation} 
 Using the above equations \ref{3dflat1} and \ref{3dflat2}, the full-sky estimator expressed in equation \ref{estimator} is re-expressed as
\begin{equation}
\begin{aligned}
    \hat{\phi}^{TT}(\kappa,\boldsymbol{\ell})&=\sum_{L,L'}\frac{\sqrt{LL'}}{2\pi}\sqrt{\frac{2\pi}{\ell}}\int_{0}^{2\pi}\frac{\mathrm{d}\varphi_{L}}{2\pi}\int_{0}^{2\pi}\frac{\mathrm{d}\varphi_{L'}}{2\pi}\int\frac{\mathrm{d}k}{2\pi}\int\frac{\mathrm{d}k'}{2\pi} \sum_{m,M,M'}(-1)^{m}i^{-m+M+M'}\begin{pmatrix} \ell & L & L' \\
-m & M & M' 
\end{pmatrix}\\& \times N_{\ell}^{\phi,(TT)}f_{\ell L L'}^{\phi,(TT)}\tilde{T}(k,\boldsymbol{L})\tilde{T}(k',\boldsymbol{L'}) e^{i(m\varphi_{\ell}-M\varphi_{L} -M'\varphi_{L'})}
\end{aligned}
\end{equation} 
 or in a reduced form
\begin{equation}
\begin{aligned}
     \hat{\phi}^{TT}(\kappa,\boldsymbol{\ell})&= \sum_{L,L'}\int_{0}^{2\pi}\frac{\mathrm{d}\varphi_{L}}{2\pi}\int_{0}^{2\pi}\frac{\mathrm{d}\varphi_{L'}}{2\pi}\int\frac{\mathrm{d}k}{2\pi}\int\frac{\mathrm{d}k'}{2\pi} F_{\boldsymbol{\ell},\boldsymbol{L},\boldsymbol{L'}}N_{\ell}^{\phi,(TT)}f_{\ell L L'}^{\phi,(TT)}\tilde{T}(k,\boldsymbol{L})\tilde{T}(k',\boldsymbol{L'}),
   \label{guessflatestimator}
\end{aligned}
\end{equation} 
 with the following abbreviation
\begin{equation}
\begin{aligned}
     F_{\boldsymbol{\ell},\boldsymbol{L},\boldsymbol{L'}}&=\frac{\sqrt{LL'}}{2\pi}\sqrt{\frac{2\pi}{\ell}}\sum_{m,M,M'}(-1)^{m}\begin{pmatrix} \ell & L & L' \\
-m & M & M' 
\end{pmatrix}i^{-m+M+M'} e^{i(m\varphi_{\ell}-M\varphi_{L} -M'\varphi_{L'})}.
\end{aligned}
\end{equation}
 The above expression of the estimator forces us to derive the term $F_{\boldsymbol{\ell},\boldsymbol{L},\boldsymbol{L'}}f_{\ell L L'}^{\phi,(TT)}$ in the flat-sky limit. Hence, we need to calculate the quantity $(F_{\boldsymbol{\ell},\boldsymbol{L'},\boldsymbol{L}})^{*}g_{\ell L' L}^{\phi,(TT)}$ in the flat-sky approximation. In this calculation, it is included the operation $(F_{\boldsymbol{\ell},\boldsymbol{L'},\boldsymbol{L}})^{*}_{s}D_{
L',\ell,L}$ and $(F_{\boldsymbol{\ell},\boldsymbol{L'},\boldsymbol{L}})^{*}_{s}D_{
L,\ell,L'} $ where $s=0$, $\pm 2$. On the other hand, the term $I_{\ell}(k,k')$ does not depend on the angular part but radially so there's no need to develop it. By developing the term $(F_{\boldsymbol{\ell},\boldsymbol{L'},\boldsymbol{L}})^{*}_{s}D_{
L',\ell,L}$, we obtain
\begin{equation}
\begin{aligned}
(F_{\boldsymbol{\ell},\boldsymbol{L'},\boldsymbol{L}})^{*}_{s}D_{
L',\ell,L}&=\frac{\sqrt{LL'}}{2\pi}\sqrt{\frac{2\pi}{\ell}}\sum_{m,M,M'}e^{i(m\varphi_{\ell}-M\varphi_{L} -M'\varphi_{L'})}(-1)^{m+M'} i^{m-M-M'}\int \mathrm{d}^{2}\boldsymbol{\hat{n}}_{s}Y_{L',-M'}^{*}(\boldsymbol{\hat{n}})\left[ \nabla_{0}Y_{\ell,-m}(\boldsymbol{\hat{n}})\right]\left[ \nabla_{s}Y_{L,M}(\boldsymbol{\hat{n}}) \right]\\&= \frac{\sqrt{LL'}}{2\pi}\sqrt{\frac{2\pi}{\ell}}\sum_{m,M,M'}e^{i(-m\varphi_{\ell}+M\varphi_{L} +M'\varphi_{L'})}(-1)^{m+M'} i^{m-M-M'}\int \mathrm{d}^{2}\boldsymbol{\hat{n}}_{s}Y_{L',-M'}(\boldsymbol{\hat{n}})\left[ \nabla_{0}Y_{\ell,-m}^{*}(\boldsymbol{\hat{n}})\right]\left[ \nabla_{s}Y_{L,M}^{*}(\boldsymbol{\hat{n}}) \right].
\end{aligned}
\end{equation}
 Making use of the following relation under the flat-sky approximation, $\ell\gg1$,
\begin{equation}
    e^{\pm s i(\varphi_{\ell}-\varphi)}e^{i\boldsymbol{\ell}\cdot\boldsymbol{\hat{n}}} \approx (\pm i)^{s}\sqrt{\frac{2\pi}{\ell}}\sum_{m}i^{m}_{\pm s}Y_{\ell,m}(\boldsymbol{\hat{n}})e^{-im\varphi_{\ell}},
    \label{approximation1}
\end{equation} 
one would arrive at
\begin{equation}
\begin{aligned}
(F_{\boldsymbol{\ell},\boldsymbol{L},\boldsymbol{L'}})^{*}_{s}D_{
L',\ell,L}=\sqrt{ LL'}\sum_{m,M,M'} \sqrt{\frac{2\pi}{\ell}}i^{m}e^{-im\varphi_{\ell}}\sqrt{\frac{2\pi}{L}}i^{-M}e^{iM\varphi_{L}}\sqrt{\frac{2\pi}{L'}}i^{-M'}e^{iM'\varphi_{L'}}\int \frac{\mathrm{d}^{2}\boldsymbol{\hat{n}}}{(2\pi)^{2}}
{}_{s}Y_{L',M'}^{*}(\boldsymbol{\hat{n}})\left[ \nabla{}_{s}Y_{L,M}^{*}(\boldsymbol{\hat{n}})\right]\left[ \nabla_{0}Y_{\ell,m}^{*}(\boldsymbol{\hat{n}}) \right].
\end{aligned}    
\end{equation} 
 Using \ref{approximation1} and assuming $\ell,L,L'\gg 1$, the previous equation becomes
\begin{equation}
    (F_{\boldsymbol{\ell},\boldsymbol{L'},\boldsymbol{L}})^{*}_{s}D_{
L',\ell,L}\simeq e^{si(\varphi_{L'}-\varphi_{L})}\boldsymbol{L}\cdot \boldsymbol{\ell}\int \frac{\mathrm{d}^{2}\boldsymbol{\hat{n}}}{(2\pi)^{2}}e^{i(\boldsymbol{\ell}-\boldsymbol{L}-\boldsymbol{L'})\cdot\boldsymbol{\hat{n}}}.
\end{equation} 
 Since we are dealing only with lensing of the brightness temperature $T$, the value of $s=0$ and also note that the delta function is given by $\delta(\boldsymbol{\ell})=\int\frac{\mathrm{d}^{2}\boldsymbol{\hat{n}}}{(2\pi)^{2}}e^{i\boldsymbol{\ell}\cdot\boldsymbol{\hat{n}}}$, the above equation reduces to
\begin{equation}
    (F_{\boldsymbol{\ell},\boldsymbol{L},\boldsymbol{L'}})^{*}_{s}D_{
L',\ell,L}\simeq \delta(\boldsymbol{L}+\boldsymbol{L'}-\boldsymbol{\ell})\boldsymbol{L}\cdot\boldsymbol{\ell}.
\end{equation} 
After making the same procedure for the other term $ (F_{\boldsymbol{\ell},\boldsymbol{L},\boldsymbol{L'}})^{*}_{s}D_{L,\ell,L'}$, we finally obtain the following expression
\begin{equation}
\begin{aligned}
    F_{\boldsymbol{\ell},\boldsymbol{L},\boldsymbol{L'}}f_{
\ell,L,L'}^{\phi,(TT)}&=\frac{\left(\overline{g}_{\ell,L,L'}^{\phi,(TT)}\right)^{*}\tilde{C}_{L'}^{TT}\tilde{C}_{L}^{TT}-\tilde{C}_{L}^{TT}\tilde{C}_{L'}^{TT}\left(\overline{g}_{\ell,L',L}^{\phi,(TT)}\right)^{*} }{\tilde{C}_{L}^{TT}\tilde{C}_{L'}^{TT}\tilde{C}_{L'}^{TT}\tilde{C}_{L}^{TT}-(\tilde{C}_{L}^{TT}\tilde{C}_{L'}^{TT})^{2}}\delta(\boldsymbol{L}+\boldsymbol{L'}-\boldsymbol{\ell}) = \overline{f}_{
\ell,L,L'}^{\phi,(TT)}\delta(\boldsymbol{L}+\boldsymbol{L'}-\boldsymbol{\ell})
\end{aligned}
\end{equation} 
 with the following abbreviation $\overline{g}_{\ell,L,L'}^{\phi,(TT)}=C_{L}^{TT}(k_{1},k_{1}',L)\boldsymbol{\ell}\cdot\boldsymbol{L}+C_{L'}^{TT}(k_{2},k_{2}',L')\boldsymbol{\ell}\cdot\boldsymbol{L'}$, where $C_{L}^{TT}(k,k',\ell)$ is the flat-sky power spectra of the brightness temperature which includes the redshift estimation $M_{\ell}$ and source distribution $Z_{\ell}$ and
\begin{equation}
\overline{f}_{
\ell,L,L'}^{\phi,(TT)}= \frac{\left(\overline{g}_{\ell,L,L'}^{\phi,(TT)}\right)^{*}\tilde{C}_{L'}^{TT}\tilde{C}_{L}^{TT}-\tilde{C}_{L}^{TT}\tilde{C}_{L'}^{TT}\left(\overline{g}_{\ell,L',L}^{\phi,(TT)}\right)^{*} }{\tilde{C}_{L}^{TT}\tilde{C}_{L'}^{TT}\tilde{C}_{L'}^{TT}\tilde{C}_{L}^{TT}-(\tilde{C}_{L}^{TT}\tilde{C}_{L'}^{TT})^{2}}. \end{equation} 

Now, we need to derive the flat-sky approximation of the noise covariance. The covariance is rewritten as
\begin{equation}
\begin{aligned}
N_{\ell}^{\phi,(i,j)}=&\sum_{m,M,M'}\sum_{m',M'',M'''}\frac{(-1)^{m+m'}\delta_{mm'}\delta_{MM''}\delta_{M'M'''}}{2\ell+1} \begin{pmatrix} \ell & L & L' \\
-m & M & M' 
\end{pmatrix}\begin{pmatrix} \ell & L & L' \\
-m' & M'' & M''' 
\end{pmatrix}\sum_{L,L'}\left( A_{\ell,L,L'}^{\phi,(i)} \right)^{*}\\& \times\left( A_{\ell,L,L'}^{\phi,(j)}\tilde{C}_{L}^{TT'}\tilde{C}_{L'}^{TT'}+ A_{\ell,L',L}^{\phi,(j)}(-1)^{\ell+L+L'}\tilde{C}_{L}^{TT'}\tilde{C}_{L'}^{T'T} \right)\\&=\sum_{m,M,M'}\sum_{m',M'',M'''}\frac{(-1)^{m+m'}}{2\ell+1}\begin{pmatrix} \ell & L & L' \\
-m & M & M' 
\end{pmatrix}\begin{pmatrix} \ell & L & L' \\
-m' & M'' & M''' 
\end{pmatrix} \int\frac{\mathrm{d}\varphi_{\ell}}{2\pi}e^{-i(m-m')\varphi}\int\frac{\mathrm{d}\varphi_{L}}{2\pi}e^{-i(M-M'')\varphi_{L}}\\& \times\int\frac{\mathrm{d}\varphi_{L'}}{2\pi}e^{-i(M'-M''')\varphi_{L'}} \sum_{L,L'}\left( A_{\ell,L,L'}^{\phi,(i)} \right)^{*}\left( A_{\ell,L,L'}^{\phi,(j)}\tilde{C}_{L}^{TT'}\tilde{C}_{L'}^{TT'}+ A_{\ell,L',L}^{\phi,(j)}(-1)^{\ell+L+L'}\tilde{C}_{L}^{TT'}\tilde{C}_{L'}^{T'T} \right),
\end{aligned}
\end{equation} 
 where we used the following relation $\delta_{M_{1},M_{2}}=\int\frac{\mathrm{d}\varphi}{2\pi}e^{-i(M_{1}-M_{2})\varphi} $. From the previous calculations, the noise covariance is given by
\begin{equation}
\begin{aligned}
N_{\ell}^{\phi,(i,j)}&=\int\frac{\mathrm{d}\varphi_{\ell}}{2\pi}\sum_{L,L'}\left(\frac{LL'}{2\pi} \right)^{-1}\int\frac{\mathrm{d}\varphi_{L}}{2\pi}\int\frac{\mathrm{d}\varphi_{L'}}{2\pi}\left( F_{\boldsymbol{\ell},\boldsymbol{L},\boldsymbol{L'}} \right)^{*}\left( F_{\boldsymbol{\ell},\boldsymbol{L},\boldsymbol{L'}} \right) \left( A_{\ell,L,L'}^{\phi,(i)} \right)^{*}\left( A_{\ell,L,L'}^{\phi,(j)}\tilde{C}_{L}^{TT'}\tilde{C}_{L'}^{TT'}+ A_{\ell,L',L}^{\phi,(j)}(-1)^{\ell+L+L'}\tilde{C}_{L}^{TT'}\tilde{C}_{L'}^{T'T} \right).
\end{aligned}    
\end{equation}
 In the flat-sky limit, the following term  is defined by $\delta(\boldsymbol{L}+\boldsymbol{L'}-\boldsymbol{\ell})\overline{A}_{\boldsymbol{\ell},\boldsymbol{L},\boldsymbol{L'}}^{\phi,(j)}\simeq F_{\boldsymbol{\ell},\boldsymbol{L},\boldsymbol{L'}}A_{\ell,L,L'}^{\phi,(j)} $. Using the earlier identity, $\delta_{\boldsymbol{0}}=\frac{1}{\pi}$, and assuming $\ell, L, L' \gg 1$, the noise covariance becomes
\begin{equation}
\begin{aligned}
\overline{N}_{\ell}^{\phi,(i,j)}&\simeq \int\frac{\mathrm{d}\varphi_{\ell}}{2\pi}\sum_{L,L'} LL' \int\frac{\mathrm{d}\varphi_{L}}{2\pi}\int\frac{\mathrm{d}\varphi_{L'}}{2\pi}\delta(\boldsymbol{L}+\boldsymbol{L'}-\boldsymbol{\ell})\left( \overline{A}_{\ell,L,L'}^{\phi,(i)} \right)^{*}\left( \overline{A}_{\ell,L,L'}^{\phi,(j)}\tilde{C}_{L}^{TT'}\tilde{C}_{L'}^{TT'}+ \overline{A}_{\ell,L',L}^{\phi,(j)}\tilde{C}_{L}^{TT'}\tilde{C}_{L'}^{T'T} \right).
\end{aligned}
\end{equation} 
 In the right-hand side of the equation, we can select two-dimensional coordinate system for the variables of integration, $\boldsymbol{L}$, $\boldsymbol{L'}$ and the one-dimensional $k$-wavenumber. Then, noise in the flat-sky limit is given by
\begin{equation}
\begin{aligned}
    \overline{N}_{\ell}^{\phi,(TT)}&=\int\frac{\mathrm{d}k}{2\pi}\int\frac{\mathrm{d}^{2}\boldsymbol{L}}{(2\pi)^{2}}\int\mathrm{d}^{2}\boldsymbol{L'}\delta(\boldsymbol{L}+\boldsymbol{L'}-\boldsymbol{\ell}) 2\left( \overline{A}_{\ell,L,L'}^{\phi,(TT)} \right)^{*}\left( \overline{A}_{\ell,L,L'}^{\phi,(TT)}C_{L}^{TT'}(k,k',\boldsymbol{\ell})C_{L'}^{TT'}(k,k',\boldsymbol{\ell})\right)
    \label{flatnoise}
\end{aligned}
\end{equation} 
 and the weight function in the flat-sky approximation becomes
\begin{equation}
    \overline{A}_{\ell,\boldsymbol{L},\boldsymbol{L'}}^{\phi,(TT)}=\overline{N}_{\ell}^{\phi,(TT)}\overline{f}_{\boldsymbol{\ell},\boldsymbol{L},\boldsymbol{L'}}^{\phi,(TT)}
    \label{flatweightfunction}
\end{equation}
 Substituting equation \ref{flatweightfunction} into equation \ref{flatnoise}, we obtain the expression for the noise spectrum of the lensing potential in the flat-sky limit:
\begin{equation}
     \overline{N}_{\ell}^{\phi,(TT)}= \left[\int\frac{\mathrm{d}k}{2\pi} \int\frac{\mathrm{d}^{2}\boldsymbol{L}}{(2\pi)^{2}}\int\mathrm{d}^{2}\boldsymbol{L'}\delta(\boldsymbol{L}+\boldsymbol{L'}-\boldsymbol{\ell})\overline{g}_{\boldsymbol{\ell},\boldsymbol{L},\boldsymbol{L'}}^{\phi,(TT)}\overline{f}_{\boldsymbol{\ell},\boldsymbol{L},\boldsymbol{L'}}^{\phi,(TT)}    \right]^{-1}.
     \label{noisespectrum}
\end{equation}
Therefore, our estimator defined previously in equation \ref{guessflatestimator} can be re-express as
\begin{equation}
\begin{aligned}
    \hat{\phi}^{TT}(\kappa,\boldsymbol{\ell})&=\int\frac{\mathrm{d}^{2}\boldsymbol{L}}{(2\pi)^{2}}\int\frac{\mathrm{d}k}{(2\pi)}\int\frac{\mathrm{d}k'}{(2\pi)}\int\mathrm{d}^{2}\boldsymbol{L'}\delta(\boldsymbol{L}+\boldsymbol{L'}-\boldsymbol{\ell})\overline{A}_{\ell,\boldsymbol{L},\boldsymbol{L'}}^{\phi,(TT)}\tilde{T}(k,\boldsymbol{L})\tilde{T}(k',\boldsymbol{L'}) \\&=\int\frac{\mathrm{d}^{2}\boldsymbol{L}}{(2\pi)^{2}}\int\frac{\mathrm{d}k}{(2\pi)}\int\frac{\mathrm{d}k'}{(2\pi)}\overline{A}_{\ell,\boldsymbol{L},\boldsymbol{L'}}^{\phi,(TT)}\tilde{T}(k,\boldsymbol{L})\tilde{T}(k',\boldsymbol{\ell}-\boldsymbol{L}).
    \label{flatestimator1}
\end{aligned}
\end{equation}
 The above estimator and noise spectrum given by
\begin{equation}
\begin{aligned}
    \hat{\phi}^{TT}(\kappa,\boldsymbol{\ell})=\int\frac{\mathrm{d}^{2}\boldsymbol{L}}{(2\pi)^{2}}\int\frac{\mathrm{d}k}{(2\pi)}\int\frac{\mathrm{d}k'}{(2\pi)}\overline{A}_{\ell,\boldsymbol{L},\boldsymbol{L'}}^{\phi,(TT)}\tilde{T}(k,\boldsymbol{L})\tilde{T}(k',\boldsymbol{\ell}-\boldsymbol{L})
\end{aligned}
\end{equation}
\begin{equation}
    \overline{N}_{\ell}^{\phi,(TT)}= \left[\int\frac{\mathrm{d}k}{2\pi} \int\frac{\mathrm{d}^{2}\boldsymbol{L}}{(2\pi)^{2}}\int\mathrm{d}^{2}\boldsymbol{L'}\delta(\boldsymbol{L}+\boldsymbol{L'}-\boldsymbol{\ell})\overline{g}_{\boldsymbol{\ell},\boldsymbol{L},\boldsymbol{L'}}^{\phi,(TT)}\overline{f}_{\boldsymbol{\ell},\boldsymbol{L},\boldsymbol{L'}}^{\phi,(TT)}    \right]^{-1}
\end{equation}
are the main results of this section. 
 
What we have done is to construct an estimator for the lensing potential in 3 dimensions $\phi_{\ell m}(k)$ using the spherical Fourier-Bessel transformation to deal with wide angle 21-cm survey data including the observational effects since 21-cm mapping experiments may require these corrections. The effect of lensing breaks homogeneity and induces correlations that can be used to reconstruct the lensing potential by a quadratic estimator proposed initially in \cite{Hu_2002}. It is not a surprise that a continuous transform is not practical numerically and a discrete equivalent is needed to handle it. A recent research \cite{10.1093/mnras/stz1781} derived a quadratic estimator for the 21-cm line intensity mapping by using a discretised SFB basis. The SFB transform can be discretised into a spherical Fourier–Bessel series in a finite volume. Indeed, more realistic surveys are constructed over spherical shells. The survey volume is treated as a spherical shell with inner radius $r_{\text{min}}$ and outer radius $r_{\text{max}}$ when modelling a transformation using the spherical Fourier– Bessel series. Hence the underlying field is sampled at discrete points, where this problem can be solved by setting some boundary conditions with the discrete nature of the cosmological survey which yields a way to estimate discrete spherical Fourier-Bessel coefficients. Such discretisation presents a weight term that helps spherical Bessel transform of a given order to be expressed as the sum of the coefficients obtained for a different order of the transform, with the appropriate weighting. In fact, \cite{10.1093/mnras/stz1781} demonstrate the construction of an idealised quadratic estimator for a 2D lensing field $\phi_{\ell m}$, in comparison of our work that extended to $\phi_{\ell m}(k)$ and includes the observational effects.  Hence, what it comes later is to compute numerically the noise reconstruction $N_{\ell}$ for the lensing potential but this is left for a future work.

\section{Numerical Results}\label{results}

\subsection{3-dimensional angular power spectrum}
\begin{figure}
	\centering\includegraphics[width=10.3cm]{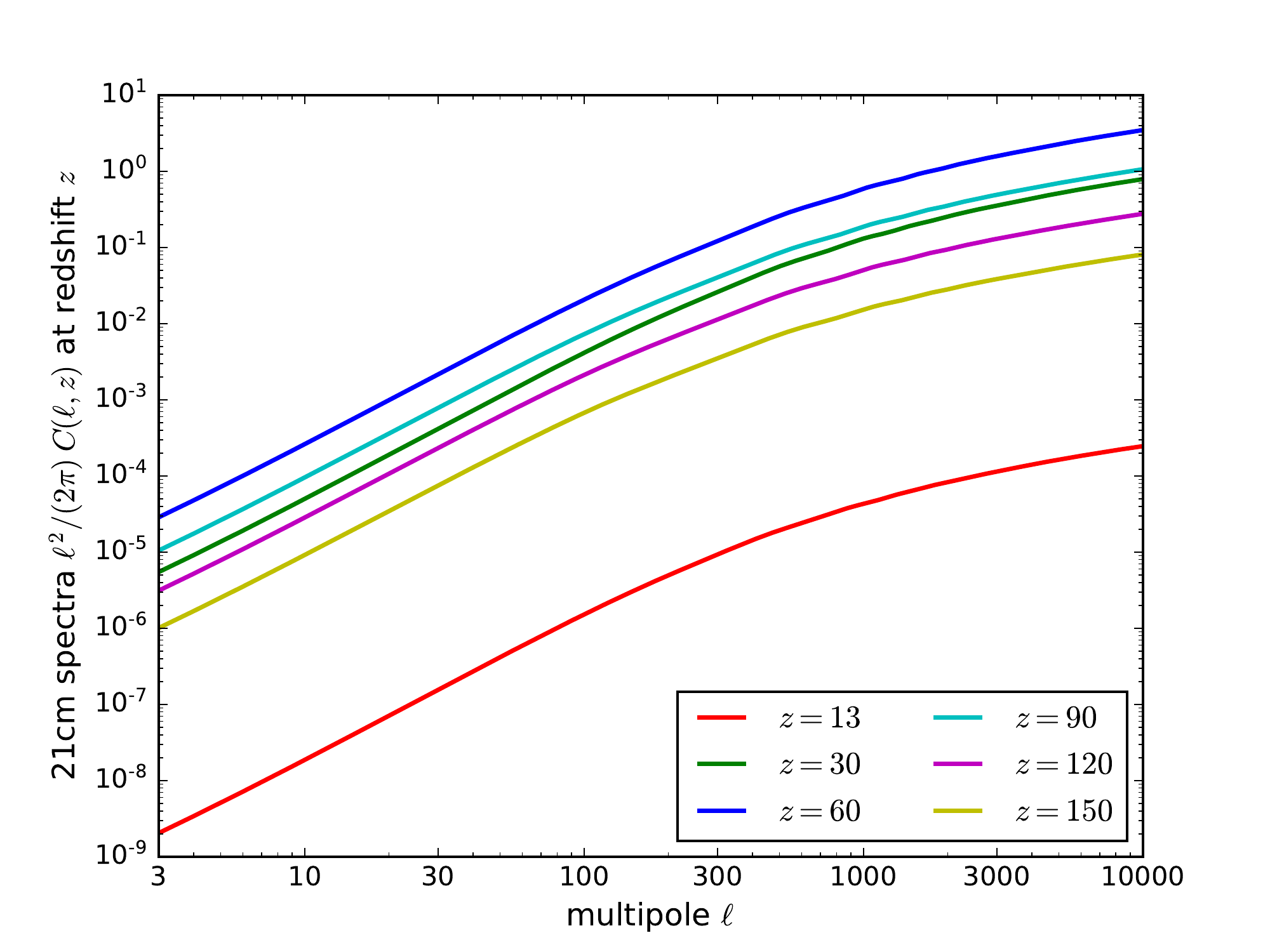}
    \caption{The 21-cm power spectrum is shown for various redshifts during the dark ages at redshifts $z=30$, $60$, $90$, $120$, $150$ and reionisation epoch at redshift $z=13$ using CAMB sources in \citet{PhysRevD.76.083005}. Here the signal is damped by the effect of baryon pressure at $\ell \geq 10^{4} $.}
    \label{3d angular power spectrum plot}
\end{figure}

The results on the angular power spectrum of the 21-cm radiation background described in equation \ref{3d21cmangularpowerspectrum} at small and large scales are exhibited for both the dark ages and the epoch of reionisation. During the dark ages, there is a redshift range $30<z<200$ where the neutral hydrogen should be visible in absorption against the CMB and free of ionisation and contamination by astrophysical objects. On the other hand, during the epoch of reionisation ($7<z<13)$, the 21-cm signal is still visible but harder to detect it since neutral hydrogen starts to be ionised due to the newly formed astrophysical structures. In Fig. \ref{3d angular power spectrum plot} we plot the 21-cm dark ages signal power spectrum at different redshifts $z=30$, $60$, $90$, $120$, $150$ and the 21-cm epoch of reionisation signal power spectrum with redshift $z=13$ for a bandwidth of 1 MHz.

\subsection{Covariance: observational and lensing effects}
\begin{figure}
	\centering\includegraphics[width=12cm]{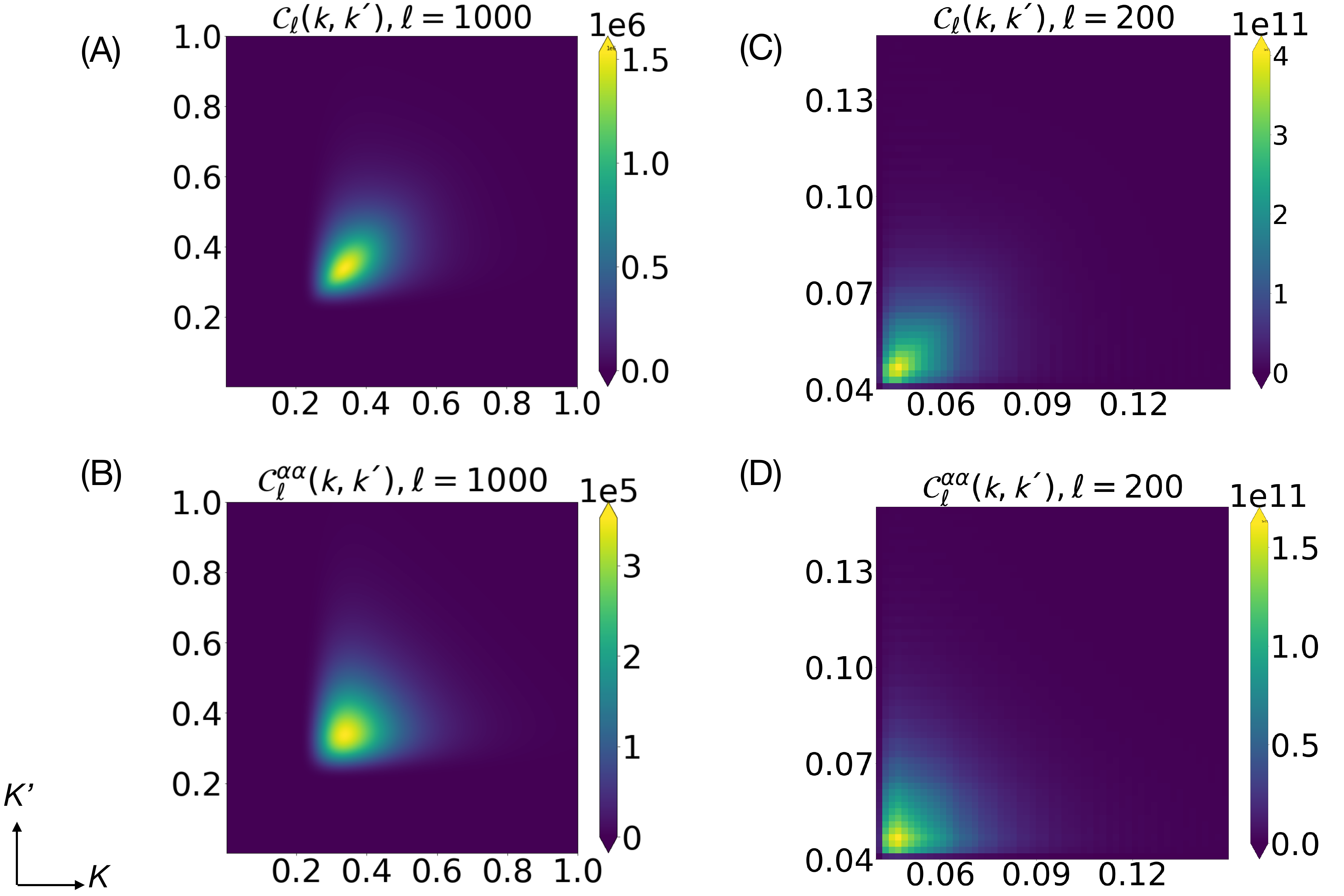}
    \caption{Figures (A) and (C) describe the covariance $C_{\ell}(k,k')$ expressed in equation \ref{covariancedeflectionangle} for multipoles $\ell=1000$ and $\ell=200$ which includes only observational effects given by $Z_{\ell}(k,k')$ and $M_{\ell}(k,k')$. On the other hand, figures (B) and (D) describe the covariance matrix of the lensing deflection angle $\boldsymbol{\hat{\alpha}}$ expressed in  equation \ref{covariancedeflectionangle} for multipole $\ell=1000$ and $\ell=200$. Figures (A) and (B) are analysed with Gaussian visibility function. As for Figures (C) and (D) are analysed with linear visibility function.  } \label{covariance100to1000}
\end{figure}

3d weak lensing analysis is based on redshift estimation of radiation sources since the exact redshift becomes difficult to measure as we go deeper and deeper into the large volume of the radiation source. Therefore, the covariance $C_{\ell}(k,k')$ encodes two observational effects that takes the redshift estimation into account. The first effect involves a comoving (radial) survey visibility function that represents the distribution in redshift of the radiation source described by the quantity $M_{\ell}(k,k')$. The second effect is the error associated to redshift estimation where the error increases as the radial (as a function of redshift $\chi_{z}$) part of the survey increases. This is described by the  conditional probability inside of the quantity $Z_{\ell}(k,k')$ that represents  the probability of estimating the redshift $z_{s}$ given the measured redshift $z_{\chi}$. So the covariance involving these two terms can be expressed by equation \ref{covarianceneutralhydrogenpowerspectrum}, where the relation between the integrals $M_{\ell}(k,k')$ and $Z_{\ell}(k,k')$ which are explained in detail in section \ref{observable effects integrals} . Since the covariance involves multiplication by rapidly oscillating spherical Bessel functions, a numerical method called Levin's collocation is applied \citep{LEVIN199695, zieser_cross-correlation_2016, PhysRevD.98.103507}. The covariance contains a characteristic integral of the form
\begin{equation}
    I[g]=\int_{z_{a}}^{z_{b}}\mathrm{d}z g(z)j_{\ell}(k_{a}\chi(z))j_{\ell}(k_{b}\chi(z)).
\end{equation} 
However, one can rewrite the above integral in terms of the comoving distance $\chi$  by using $\mathrm{d}z=\mathrm{d}\chi E[z(\chi)]$. It leads to
\begin{equation}
I[g]=\frac{1}{\chi_{H}}\int_{\chi(z_{1})}^{\chi(z_{2})}\mathrm{d}\chi E[z(\chi)]g[z(\chi)]j_{\ell}(k_{a}\chi(z))j_{\ell}(k_{b}\chi(z)).
\end{equation}
  We can compute equation \ref{covarianceneutralhydrogenpowerspectrum} with the conditional probability distribution for the estimated redshift $z_{s}$ given the true redshift $z_{\chi}$ to be a Gaussian given in equation \ref{conditionalprobability} with a redshift-dependent dispersion $\sigma(z)=\sigma_{z}(1+z)$ and 3-dimensional source distribution given by a Gaussian and linear form
\begin{equation}
    \tilde{W}(z)\mathrm{d}z\propto \text{e}^{-\left(z/z_{m}\right)^{2}} \mathrm{d}z, \quad  \tilde{W}(z)\mathrm{d}z\propto \left(\frac{z}{z_{m}}\right) \mathrm{d}z
    \label{source distribution}
\end{equation} 
 The results of the computation of the covariance without and with lensing for the above visibility functions are depicted in Fig. \ref{covariance100to1000}. The covariance computed for both visibility functions has the characteristic that the value of $C_{\ell}$ falls off rapidly from the diagonal $k=k'$ because of the rapid oscillations of the multiple Bessel integrals. This means that the correlation  with $k\neq k'$ is weak outside of the diagonal $k=k'$.

 Secondly, we assume that the unlensed 21-cm radiation background is a statistically homogeneous Gaussian random field with zero mean. When we include the lensing term, it can be thought of as introducing into the 21-cm radiation background small contributions of a non-Gaussian field. As a result, the lensing term introduces off-diagonal elements into the radiation background covariance so that it provides us ways of extracting lensing signal from the observed radiation background. In this case, we introduce the effect of lensing into the previous covariance expressed in equation \ref{covarianceneutralhydrogenpowerspectrum}. Unlike the cosmic shear covariance, which is the second partial derivative of the gravitational potential, $\gamma=\partial^{2}\Phi$, the lensing deflection angle is just the first partial derivative of the gravitational potential, $ \alpha = \partial\Phi$, with lensing kernel
\begin{equation}
\begin{aligned}
   \eta_{\ell }^{\alpha}(k,k')&=\frac{4}{\pi} \int_{0}^{\infty}\mathrm{d}\chi \chi^{2}j_{\ell}(k\chi)\int_{0}^{\chi}\mathrm{d}\chi' \left[ \frac{\chi-\chi'}{\chi} \right] j_{\ell}(k'\chi')\frac{D[a(\chi')]}{a(\chi')}.
    \label{lensingterm}
\end{aligned}
\end{equation}
 We look not for the distortion but the differential shifting of the points projected on the sky: By introducing the lensing term expressed in equation \ref{lensingterm}, which includes not only the integrals corresponding to the redshift distribution $Z_{\ell}(k,k')$ and the source distribution $M_{\ell}(k,k')$ but also the lensing term which couples different $\ell$-modes breaking statistical homogeneity. The covariance expressed in equation \ref{covariancedeflectionangle} has been computed and presented in Fig.~\ref{covariance100to1000} (B, D) for different values of $\ell$ with conditional probability and visibility functions expressed in equations \ref{conditionalprobability} and \ref{source distribution}, respectively. We compare the shape of the 3-dimensional covariance matrices depicted in Figs.~\ref{covariance100to1000} $(A)$ and $(B)$ and Figs.~\ref{covariance100to1000} in $(C)$ and $(D)$. We show from top to bottom the covariance without the lensing term for multipoles $\ell=200$ and $\ell=1000$. As we observe, the covariance matrix shows small correlations in the plane $(k,k')$ for low $\ell$. The small correlation at low $\ell$ is explained by the high oscillatory nature of spherical Bessel functions. Besides, the highest values of the covariance are found diagonally along the $(k,k')$ plane. On the other hand, the covariance in Fig.~\ref{covariance100to1000} $(B)$ and $(D)$ exhibits slightly more correlations of $k,k'$ pairs. This is because lensing introduces non-diagonal terms in the covariance.

\subsection{Variation of the visibility function}
\begin{figure}
	\centering\includegraphics[width=12cm]{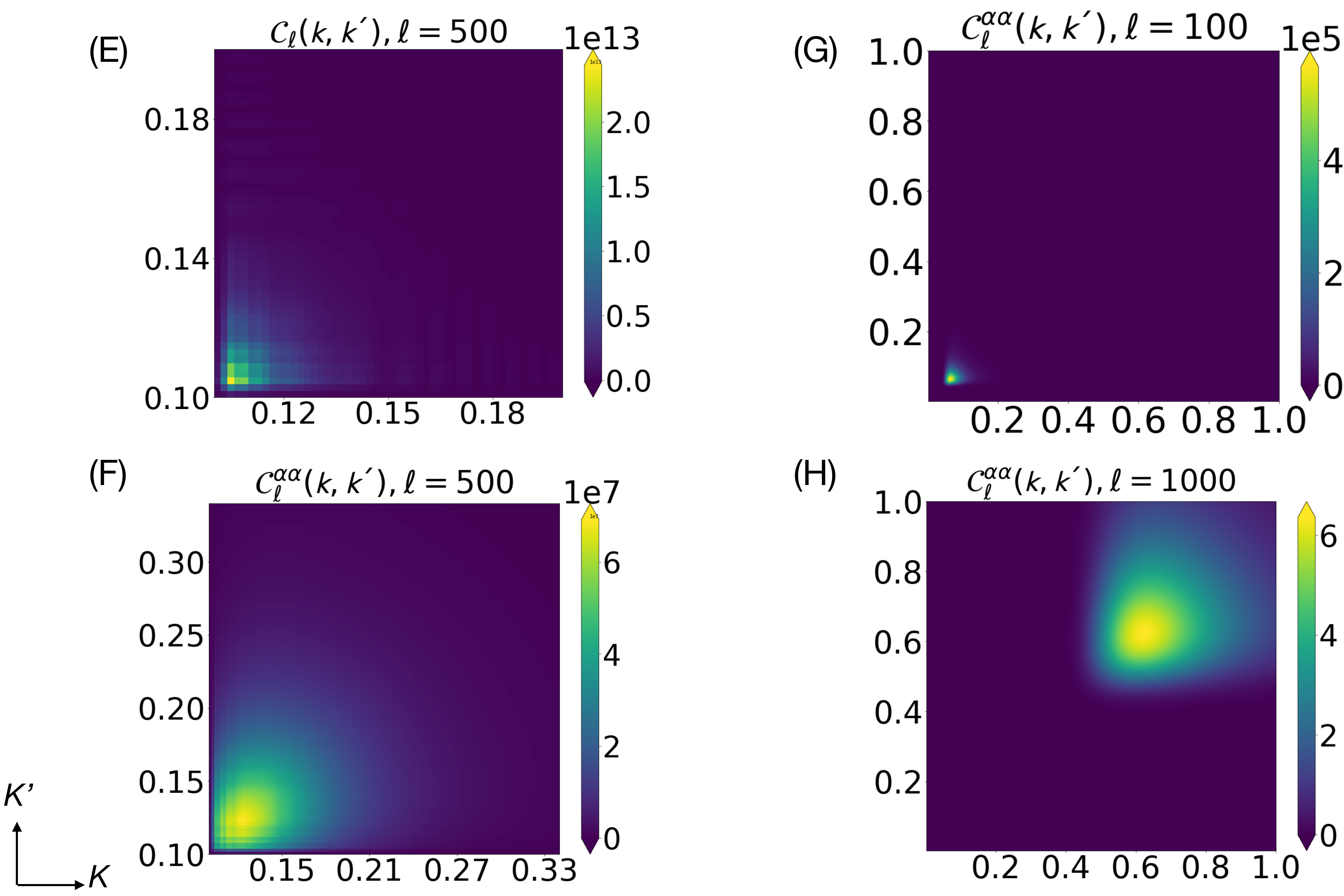}
    \caption{Figures (E)-(F) describe the covariance without and with lensing effect for multipole $\ell=500$, respectively. In this case, the lensing effect is highly notorious with the first visibility function given in equation \ref{FERMIDIRACSOURCEDISTRIBUTION}. On the other hand, figures (G)-(H) show the covariance with lensing effect for multipoles $\ell=100$ and $\ell=1000$ with flattened Gaussian source distribution indicated in the second equation given in \ref{FERMIDIRACSOURCEDISTRIBUTION}.}\label{variationofreionisation}
\end{figure}

As we mentioned before, there are two observational effects playing a key role for the lensing effect: the source distribution and the redshift estimation, in our case specifically the visibility function and the receiver bandwidth, translated into a radial brightness profile and a redshift resolution, respectively. The influence of different visibilites is depicted in Fig. \ref{covariance100to1000} in the quanity $C_{\ell}(k,k')$. Another possible choice for the visibility approximated to the evolution of the mean neutral hydrogen fraction $\overline{x}_{HI}$ is given by 
\begin{equation}
    \tilde{W}(z)\mathrm{d}z\propto\frac{1}{1+\text{exp}\left(z/z_{m}\right)}\mathrm{d}z \quad \text{and}\quad \tilde{W}(z)\mathrm{d}z\propto n\left(\text{exp}\left[-\left(\frac{z}{z_{m}} \right)^{n}\right]\right)\mathrm{d}z, 
    \label{FERMIDIRACSOURCEDISTRIBUTION}
\end{equation}
 where we set $z_{m}=0.9$ for a deeper survey. Such a visibility function is implemented in the lensed covariance expressed in equation \ref{covariancedeflectionangle}. The results are depicted in Fig. \ref{variationofreionisation} $(E)$ for the unlensed covariance and Fig.\ref{variationofreionisation} $(F)$ for the lensed covariance. The correlation between $(k,k')$ values of non-diagonal terms becomes notorious as the lensing effect is introduced at multipole $\ell=500$. Furthermore, we also compute the lensed covariance with a visibility function in the form of a flattened Gaussian, with n=3.  The results of the lensed covariance with this flattened Gaussian  function are depicted in Fig. \ref{variationofreionisation} $(G)$ and $(H)$.

\section{Summary}\label{conclusions}
It is universally recognised that the 21-cm radiation background coming from the hyperfine transition of neutral hydrogen provides cosmological data from the reionisation and dark period. Observed fluctuations of the 21-cm signal from high redshifts should result from a statistically homogeneous and isotropic Gaussian random process. If, however, this observed image has been distorted by weak gravitational lensing from the large scale structures, a breaking of statistical homogeneity is introduced in the lensed field and consequently, the covariance becomes non-diagonal. Hence, the effects of weak lensing of 21-cm radiation background are almost the same as those on CMB fluctuations and can thus be analysed with a similar formalism developed for lensing of the CMB. The 21-cm signal of the neutral hydrogen has emerged as a cosmological probe to explore deep regions of the sky due to its emission at different values of redshift. This sea of radiation probes a long period of the cosmic history from the decoupling up to the reionisation epoch. To analyse the 21-cm signal from both angular and radial regions of a such wide area observations, the spherical Fourier-Bessel decomposition is introduced as a natural basis, formed by the usual spherical harmonics $Y_{\ell m}(\boldsymbol{\hat{n}})$ and Bessel functions $j_{\ell}(k\chi)$. The analysis of the 21-cm radiation background resembles to that of the CMB. The 3d power spectrum of the brightness temperature is derived in the SFB basis including observational effects. The resulting covariance of the 21-cm radiation background is an integral over mixing matrices composed by the redshift distribution $M_{\ell}$, redshift error $Z_{\ell}$ and lensing $\eta_{\ell}$. The matrices are sources of correlations in $k$, leading to inhomogeneities. The computation of the covariance is however numerically difficult and requires a fast evaluation of highly oscillatory integrals, which is solved by the technique called Levin collocation. The variation of the visibility function in the total covariance has a strong influence since it probes large areas of the sky, may be limited to a subset of the full-sky information. The lensing effect introduces off-diagonal elements with different $\ell$ or $m$, hence it can be used to reconstruct the lensing potential by a quadratic estimator and hence noise lensing reconstruction.

\section*{Acknowledgements}
We would like to thank to M. Bartelmann for comments during the development of this research project.

\section*{Data Availability}
The data underlying this article will be shared on reasonable request to the corresponding author.

\bibliographystyle{mnras}
\bibliography{references} 

\appendix

\bsp	
\label{lastpage}
\end{document}